\documentclass[floatfix, prb, aps,amsmath,amssymb,  twocolumn,superscriptaddress,10pt,showkeys, author-year]{revtex4-2}

\usepackage[usenames,dvipsnames]{xcolor}
\usepackage{hyperref}
\usepackage{orcidlink}
\usepackage{placeins}
\usepackage{rotating}
\usepackage[export]{adjustbox}
\usepackage{soul}
\hypersetup{
colorlinks=true,
citecolor=NavyBlue,
linkcolor=NavyBlue,
filecolor=NavyBlue,
urlcolor=NavyBlue
}

\usepackage{amsmath}
\usepackage{graphicx}

\usepackage{comment}
\usepackage[normalem]{ulem}

\begin{document}
\title{Self-assembled clusters of mutually repelling particles in confinement}

\author{P. D. S. de Lima\orcidlink{0000-0002-7353-536X}}
\affiliation{Departamento de Física Teórica e Experimental, \\Universidade Federal do Rio Grande do Norte, 59078-970 Natal-RN, Brazil}
\author{R. De La Cour}
\affiliation{School of Physics, Trinity College Dublin, Dublin 2, Ireland}%
\affiliation{School of Physics, Technological University Dublin, Dublin 7, Ireland}%
\author{K. Gaff}
\affiliation{School of Physics, Trinity College Dublin, Dublin 2, Ireland}%
\author{J. M. de Araújo\orcidlink{0000-0001-8462-4280}}
\affiliation{Departamento de Física Teórica e Experimental, \\Universidade Federal do Rio Grande do Norte, 59078-970 Natal-RN, Brazil}
\author{S. J. Cox\orcidlink{0000-0001-6129-3394}}
\affiliation{Department of Mathematics, Aberystwyth University, Aberystwyth, Ceredigion, SY23 3BZ, Wales,  U.K.}
\author{M. S. Ferreira\orcidlink{0000-0002-0856-9811}}
\affiliation{School of Physics, Trinity College Dublin, Dublin 2, Ireland}%
\affiliation{Centre for Research on Adaptive Nanostructures and Nanodevices (CRANN) \& Advanced Materials and Bioengineering Research (AMBER) Centre, Trinity College Dublin, Dublin 2, Ireland}
\author{S. Hutzler\orcidlink{0000-0003-0743-1252}}
\affiliation{School of Physics, Trinity College Dublin, Dublin 2, Ireland}%

\date{\today}

\begin{abstract}
Mutually repelling particles form spontaneously ordered clusters when forced into confinement. The clusters may adopt similar spatial arrangements even if the underlying particle interactions are contrastingly different. Here we demonstrate with both simulations and experiments that it is possible to induce particles of very different types to self-assemble into the same ordered geometric structure by simply regulating the ratio between the repulsion and confining forces. This is the case for both long- and short-ranged potentials. This property is initially explored in systems with two-dimensional (2D) circular symmetry and subsequently demonstrated to be valid throughout the transition to one-dimensional (1D) structures through continuous elliptical deformations of the confining field. We argue that this feature can be utilized to manipulate the spatial structure of confined particles, thereby paving the way for the design of clusters with specific functionalities.

\end{abstract}

\maketitle

\section{Introduction}

Understanding the mechanisms that govern the formation and evolution of particle clusters is of fundamental interest in scientific areas as diverse as statistical physics~\cite{schehr2019, kundu2022}, condensed matter physics~\cite{frenkel2002, cao2003, andrey2018}, and astrophysics, to name a few. The composing particles may be magnetic~\cite{kwok2009, wolff2015, andrey2018, deLimaEtal2024} and/or ionic \cite{PartnerEtal2013}, thermally and/or optically active~\cite{optical_active_cluster}, and may range in size from the nanometer up to very large scales \cite{PhysRevLett.108.268303, PhysRevX.14.041061}. The breadth of particle types and sizes of these clusters means that they may be controlled through different types of force fields, suggesting diverse applications~\cite{ocean-cleaning, nano10071318, 9076325, Liao2023}. The coexistence of two competing force fields acting on the particles is what drives their spontaneous clustering: one that traps them within a restricted area of space and a pairwise repulsion that prevents them from coalescing into a high-density region.

%\begin{figure}
%\centering    
%\includegraphics[width=0.5\linewidth, angle=90]{figures/Hard sphere data/HS files/e = 0.87/N=7.png}
%\includegraphics[width=0.8\linewidth, angle=180]{figures/Magnet exp/Individual magnet images without grid lines/e=0.87/N=7.png}
%\includegraphics[width=0.5\linewidth, angle=270]{figures/bubble-experiments/Individual foam photos/e=0.87/N=7.2.jpeg}
%\caption{Assemblies of particles governed by very different physical interactions may display resembling equilibrium arrangements. All three images correspond to experimental realisations of $N=7$ objects confined within an ellipse of eccentricity $\epsilon=0.87$. (a) close-packed hard spheres; (b) Magnetically active particles in the form of floating dipoles; (c) soap bubbles. Details about the experimental setups can be found in Section \ref{s:experiments}. 
%\remSH{I need to double-check the hard sphere packing.}
%}
%\label{fig:similarities}
%\end{figure}

\begin{table*}
%\begin{table*}[hbt!]
    \centering
    \begin{tabular}{|c|c|c|c|c|c|}
\hline
$\epsilon$      & 0 & 0.44 & 0.66 & 0.87 & 0.95 \\
\hline
{\bf N=5} &&&&& \\
\begin{turn}{90} hard spheres \end{turn} & 
\includegraphics[width=0.13\linewidth, angle = 0]{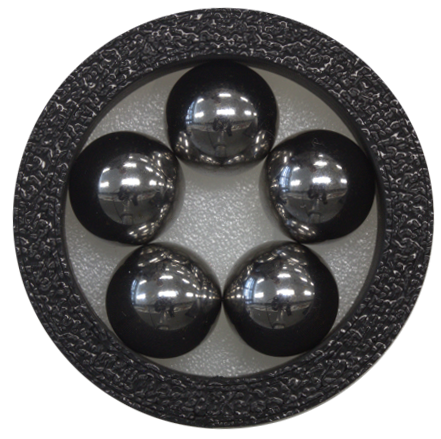}
& 
\includegraphics[width=0.1372\linewidth, angle = 0]{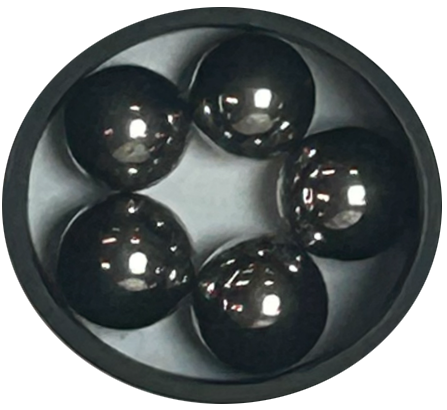}
&
\includegraphics[width=0.150\linewidth, angle = 0]{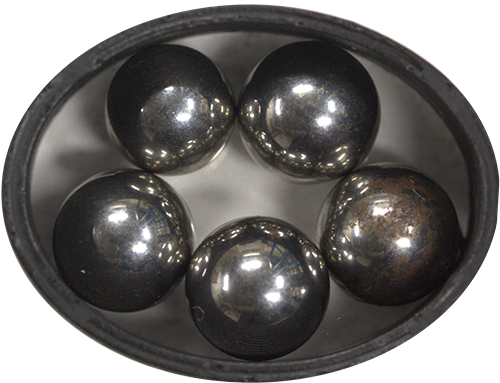}
& 
\includegraphics[width=0.1851\linewidth, angle = 0]{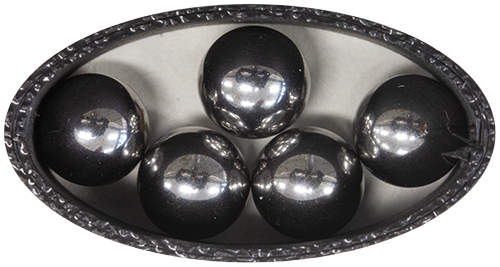}
&
\includegraphics[width=0.105\linewidth, angle = 90]{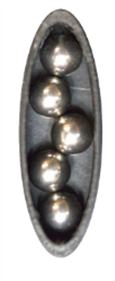}\\
\hline

\begin{turn}{90} \shortstack{floating\\magnets} \end{turn} &
\includegraphics[width=0.13\linewidth, angle = 0]{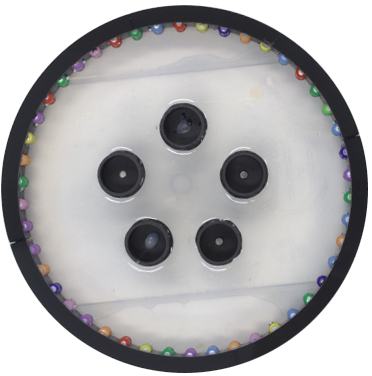}
& 
\includegraphics[width=0.137\linewidth, angle = 0]{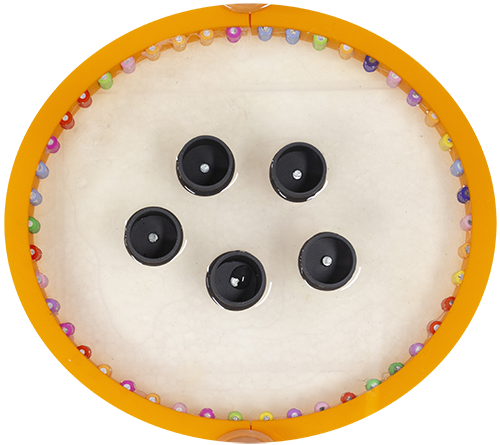}
&
\includegraphics[width=0.150\linewidth, angle = 0]{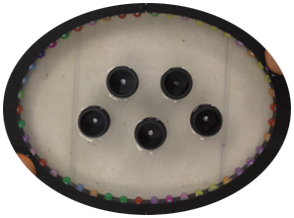}
&
\includegraphics[width=0.185\linewidth, angle = 0]{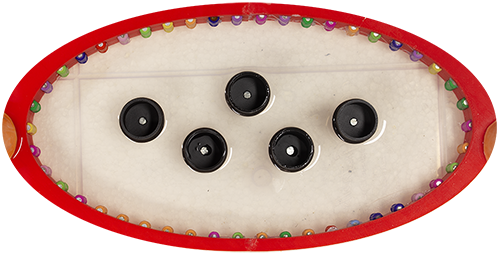}
&
\includegraphics[width=0.233\linewidth, angle = 0]{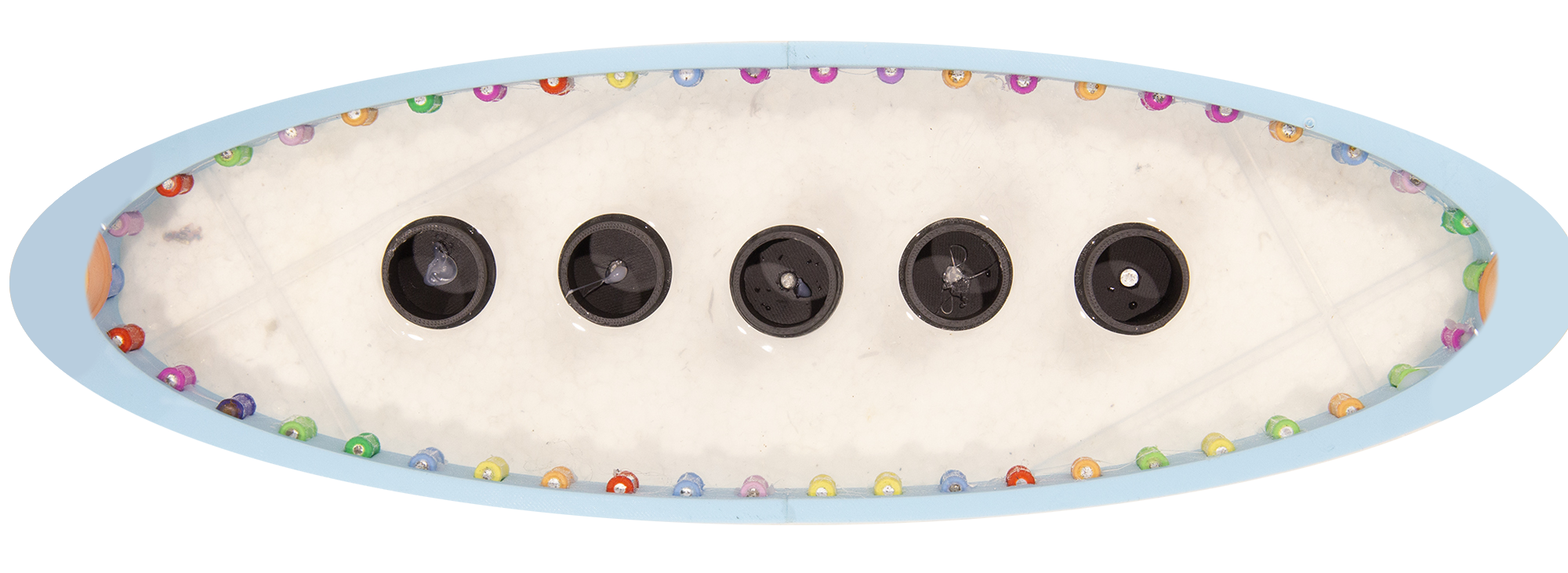}\\
% &  &  & &  \includegraphics[width=0.2\textwidth, angle=0]{figures/Magnet exp/Black floaters/e=0.87 (2)/N=5.2.png}&\\
\hline

\begin{turn}{90} bubbles \end{turn} & 
\includegraphics[width=0.13\linewidth, angle = 0]{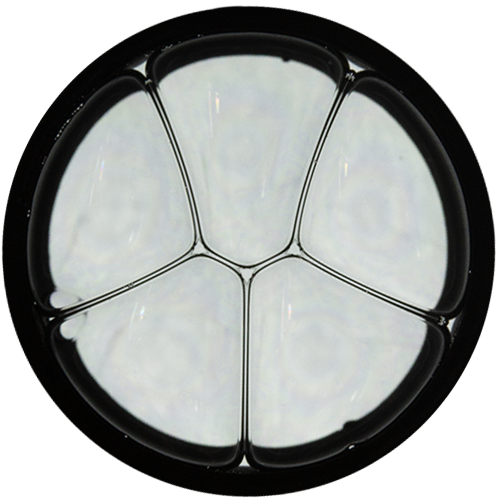}
& 
\includegraphics[width=0.137\linewidth, angle = 0]{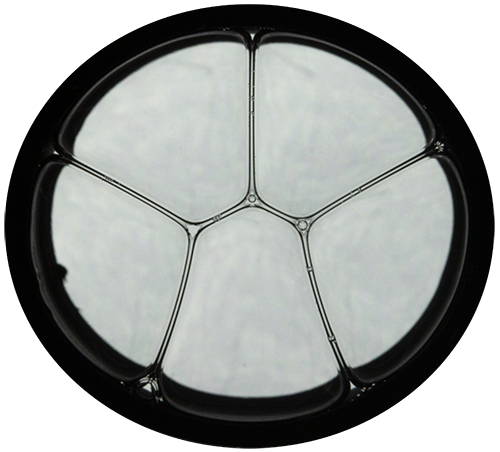}
& 
\includegraphics[width=0.150\linewidth, angle = 0]{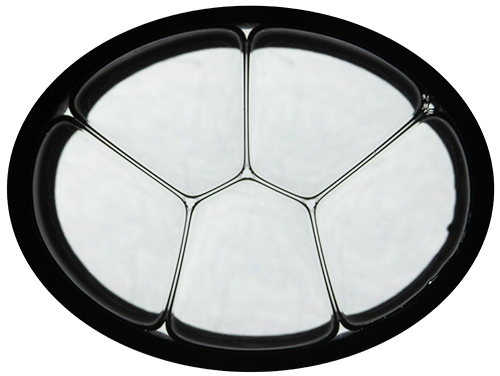}
& 
\includegraphics[width=0.185\linewidth, angle = 0]{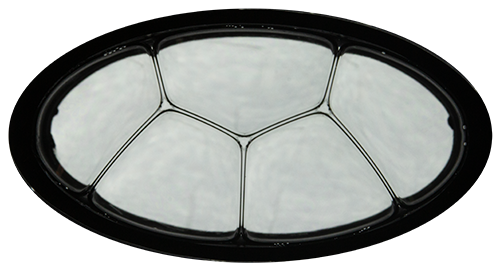}
&
\includegraphics[width=0.233\linewidth, angle = 0]{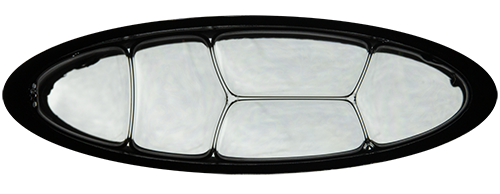}
\\

 %&  &  & & & \includegraphics[width=0.1\textwidth, angle = 90]{figures/bubble-experiments/Individual foam photos/e=0.95/N=5.2.jpg}\\

\hline\hline

{\bf N=10} &&&&& \\
\begin{turn}{90} hard spheres \end{turn}  & 
\includegraphics[width=0.13\linewidth, angle = 0]{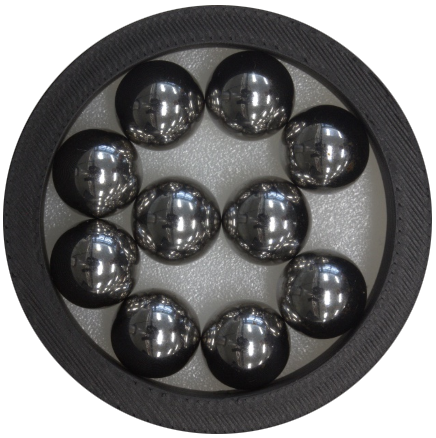}
& 
\includegraphics[width=0.1372\linewidth, angle = 0]{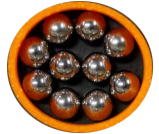}
& 
\includegraphics[width=0.1500\linewidth, angle = 0]{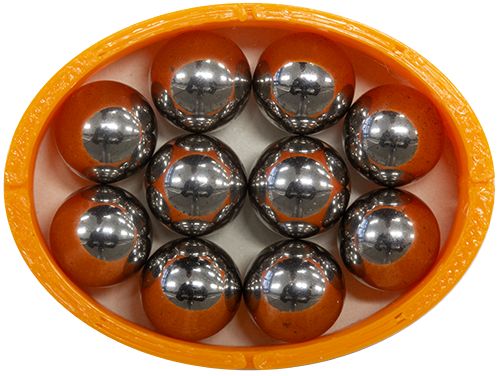}
& 
\includegraphics[width=0.1851\linewidth, angle = 0]{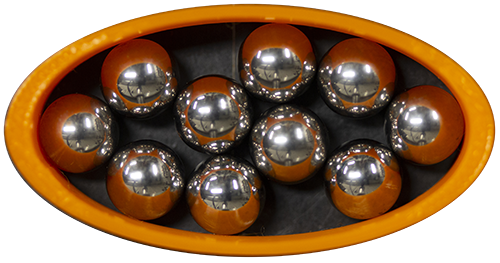}
& 
\includegraphics[width=0.2327\linewidth, angle = 0]{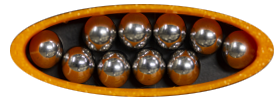}\\
\hline

\begin{turn}{90} \shortstack{floating\\magnets} \end{turn}     & 
\includegraphics[width=0.13\linewidth, angle = 0]{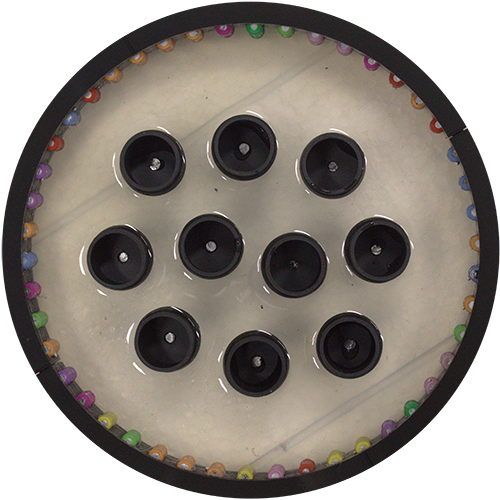}
& 
\includegraphics[width=0.1372\linewidth, angle = 0]{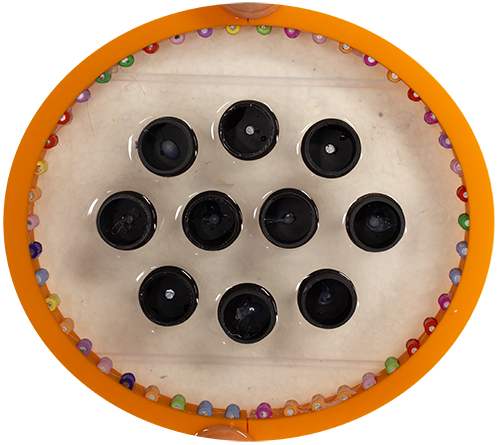}
& 
\includegraphics[width=0.1500\linewidth, angle = 0]{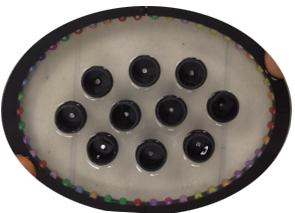}
& 
\includegraphics[width=0.1851\linewidth, angle = 0]{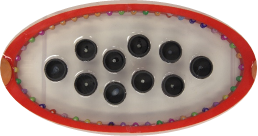}
& 
\includegraphics[width=0.2327\linewidth, angle = 0]{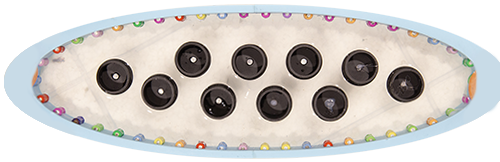}
\\
\hline

\begin{turn}{90} bubbles \end{turn}     &
\includegraphics[width= 0.13\linewidth, angle = 0]{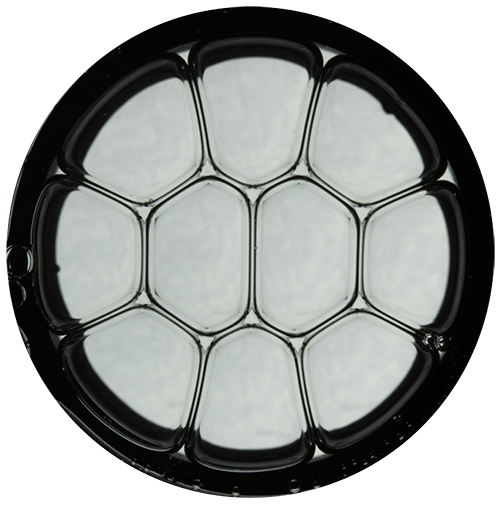}
&  
\includegraphics[width=0.1372\linewidth, angle = 0]{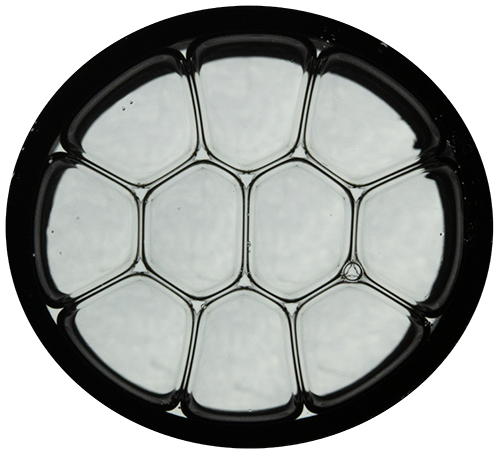}
& 
\includegraphics[width=0.1500\linewidth, angle = 0]{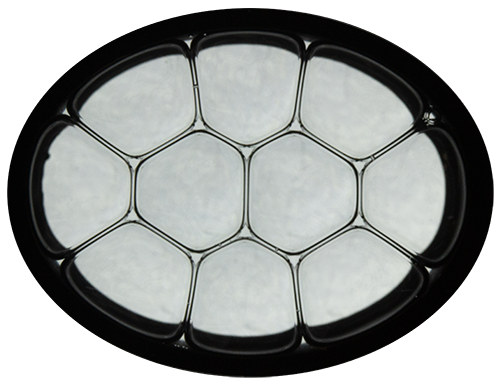}
& 
\includegraphics[width=0.1851\linewidth, angle = 0]{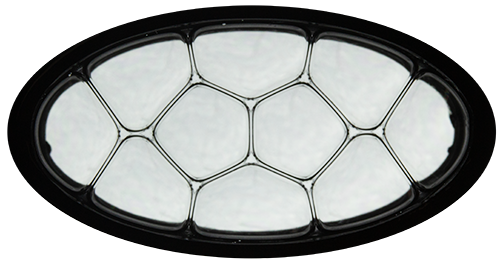}
& 
\includegraphics[width=0.2327\linewidth, angle = 0]{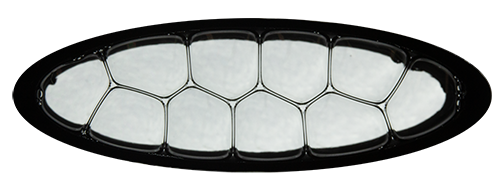}
\\
%&  &&&\includegraphics[width=0.2\textwidth, angle = 0]{figures/bubble-experiments/Individual foam photos/e=0.87/N=10.4.png}&\\
\hline

\end{tabular}
\caption{
Interacting particles in confinement arrange themselves in similar configurations, despite the very different physical nature of their interactions.
    The table shows photographs of $N=5$ and 10 close-packed hard spheres,  magnetically active particles in the form of floating dipoles, and soap bubbles. All are confined within ellipses, for five different values of eccentricity, $\epsilon=0, 0.44, 0.66, 0.87$ and $0.95$. (Details about the respective experimental set-ups are provided in Section \ref{s:experiments}.) 
    Alternative arrangements exist (see Appendix \ref{a:further-structures}); here we show only those of minimal energy, as identified in computer simulations (Section \ref{s:simulations}). The hard sphere packings for  $\epsilon=0$ (circular confinement) correspond to conjectured densest circle packings, as obtained from simulations~\cite{packomania}, also $N=5$, $\epsilon = 0.87$ and $N=10$, $\epsilon = 0.66$~\cite{Friedman}.
 }
    \label{t:comparison}
\end{table*}

%ADD A SHORT PARAGRAPH PROVIDING SEVERAL EXAMPLES OF SUCH CLUSTERS WITH DIFFERENT POTENTIALS (e.g. PRL 123, 100603 (2019), PRL 128, 170603 (2022), \cite{jimaging7050082}, and many more).

%Despite the differences in the interaction potentials, the examples above illustrate that there are common features in all these clusters. Furthermore, the%
Previous studies of such clusters (sometimes referred to as `classical Wigner clusters' \cite{RadzvilavivciusEtal2011}) %\remSH{of such `classical 2D Wigner clusters'}
have reported that particles may self-assemble into similar spatial arrangements when their underlying interactions are not too different. 
Numerical results were obtained for particles placed in a harmonic (circular) confining potential and interacting via a Coulomb potential~\cite{BoltonRoessler1993, bedanovpeeters94, KongEtal2002}, but also for elliptic confinement, with particles interacting via Coulomb \cite{RancovaEtal2011} or a Yukawa potential \cite{CandidoEtal1998,lailin99}, or with a logarithmic force law \cite{ApolinarioEtal2005}. 
In Ref. \cite{lailin99}, a parameter could be varied to approximate arrangements of hard spheres in confinement; Ref. \cite{packomania} is an online database of conjectured optimal sphere packings in circular confinement, obtained computationally, for a large range of sphere numbers. Experimental studies include arrangements of magnetic particles~\cite{Mayer1878a, Mayer1878b, RiverosEtal2004, NemoianuEtal2022, deLimaEtal2024} or those interacting via a Coulomb potential~\cite{SaintJeanEtal2001,SaintJeanGuthmann2002}.

%\remSH{I have only very briefly mentioned the simulations of others. They might require more scrutiny?}

In this article, we argue that the conditions that interaction potentials must obey to generate clusters with the same equilibrium structure might be much less restrictive than previously thought. As a result, it is possible to identify systems that, at first glance, have very little in common with one another and yet display similar geometrical layouts. In fact, here we bring together examples of electrically charged particles, magnetic dipoles, hard spheres, and bubbles in a liquid foam, all under a common framework. We show that a simple model based on the repulsion-confinement ratio can describe cluster formation in widely diverse systems. This suggests a strong degree of universality of self-assembled  particle clusters in general, as seen in  
%Fig. \ref{fig:similarities} 
Table \ref{t:comparison}
where three of the aforementioned systems display similar geometrical arrangements, even though the nature and the range of the interaction between the parts are totally different.

Furthermore, given the expected sensitivity of the spatial arrangement of particles to system dimensionality, we focus on the evolution as particles initially confined to a two-dimensional environment are deformed towards a one-dimensional structure. For simplicity, we constrain the clusters to elliptical geometries, where a circle (of eccentricity $\epsilon=0$) can be continuously transformed into a line as $\epsilon\rightarrow1$.

Published results regarding particles in elliptical confinement appear to be limited, motivating our experiments, depicted in Table \ref{t:comparison}, and the simulations that follow.
Saint Jean and Guthmann performed experiments with electrically charged stainless steel balls placed within a charged elliptic confinement \cite{SaintJeanGuthmann2002}. The structures they found are consistent with our results, as will be discussed in Section \ref{s:discussion}. Systematic computational results for particles confined within an ellipse also appear to be limited, due to the much more demanding computation in determining circle-ellipse overlaps. For the hard-sphere problem, Birgin {\em et al.} describe a nonlinear optimization procedure which computes dense packings of discs within an ellipse \cite{BirginEtal2013}. (However, their main interest relates to algorithm efficiency.) Amore {\em et al.}~\cite{AmoreEtal2023} compute the variation in packing density with the shape of the ellipse for a limited range of numbers of particles. 

%\st{In contrast to prior simulation-based studies [CITE], we present supporting experimental evidence for all considered cases.} \remSH{Also others did expts and theory.} %{\remMF{Moreover, we are able to determine the precise eccentricity required for each experiment to achieve a desired equilibrium configuration. This constitutes a significant advancement towards the rational design of particle clusters with tailored functionalities. TBC.}}

The structure of the manuscript is as follows. The next section introduces the model that will guide us throughout the manuscript to describe the equilibrium configurations of the particles in all the clusters. In addition, we demonstrate that under certain conditions it is possible for a cluster to retain its equilibrium configuration even if the interaction between the composing particles is drastically modified. Sections \ref{s:experiments} and \ref{s:simulations} present a summary of experimental and numerical results, respectively, obtained for a number of different systems. The particle arrangements themselves are discussed in section \ref{s:comparison}.  Finally, Section \ref{s:discussion} addresses the potential impact of our findings and how they can be used in real applications.

\section {Guiding model}
\label{energetics}

To illustrate our theoretical argument, we start by considering a confining region of size $R$ and finding the exact equilibrium configuration of a cluster composed of particles interacting through the pairwise central potential energy
\begin{equation}
   u(r) = u_0(r/R)^{-\alpha}\,,
   \label{eq:pairwise}
\end{equation} 
%\begin{equation}
%   u(r) = A \, \left(\frac{r}{R}\right)^{-\alpha}\,\,,
%   \label{eq:pairwise}
%\end{equation} 
where $r$ is the distance between particles and $\alpha$ is a dimensionless exponent. The constant $u_0$ has dimensions of energy, but its exact form depends on the nature of the interaction. For example, it is proportional to the product of charges in the case of Coulomb interaction and to the product of magnetic moments in the case of interacting magnetic dipoles. The exponent $\alpha$ defines not only the type of particle interaction described by Eq. \eqref{eq:pairwise} but also gives an indication of its effective range. It is instructive to separate the different confining symmetries between circular and elliptic since the former provides a much simpler illustrative picture. 

\subsection{Circular confinement}

For particles contained within a circular confinement of radius $R$ we introduce the dimensionless distance $\rho$ as $\rho = r/R$ we can then rewrite Eq. \eqref{eq:pairwise} in the dimensionless form
\begin{equation}
   U(\rho) = u(r) / u_0 = \rho^{-\alpha}\,\,.
   \label{eq:pairwise-nondim-sh}
\end{equation} 
%(N.B. not much is said about the elliptical coordinates in the current script. Length normalisation would be according to $\rho=r/\sqrt{ab}$, with $a,b$ semi-major and semi-minor axis.)

For simplicity, it is convenient to have the Coulomb interaction as a reference, {\it i.e.} $\alpha=1$. Let us then consider a cluster composed of $N$ charged particles in the presence of a circularly confining force field. We choose the latter to be generated by $M$ charged particles whose charges are smaller than those of the cluster by a factor $\lambda$, where  $\lambda$ is a dimensionless positive constant. It is worth recalling that the cluster formation results from a competition between confinement and repulsion, and the parameter $\lambda$ plays a key role in regulating the balance between the two. 

\begin{figure*}
    \centering
    \includegraphics[width=\columnwidth]{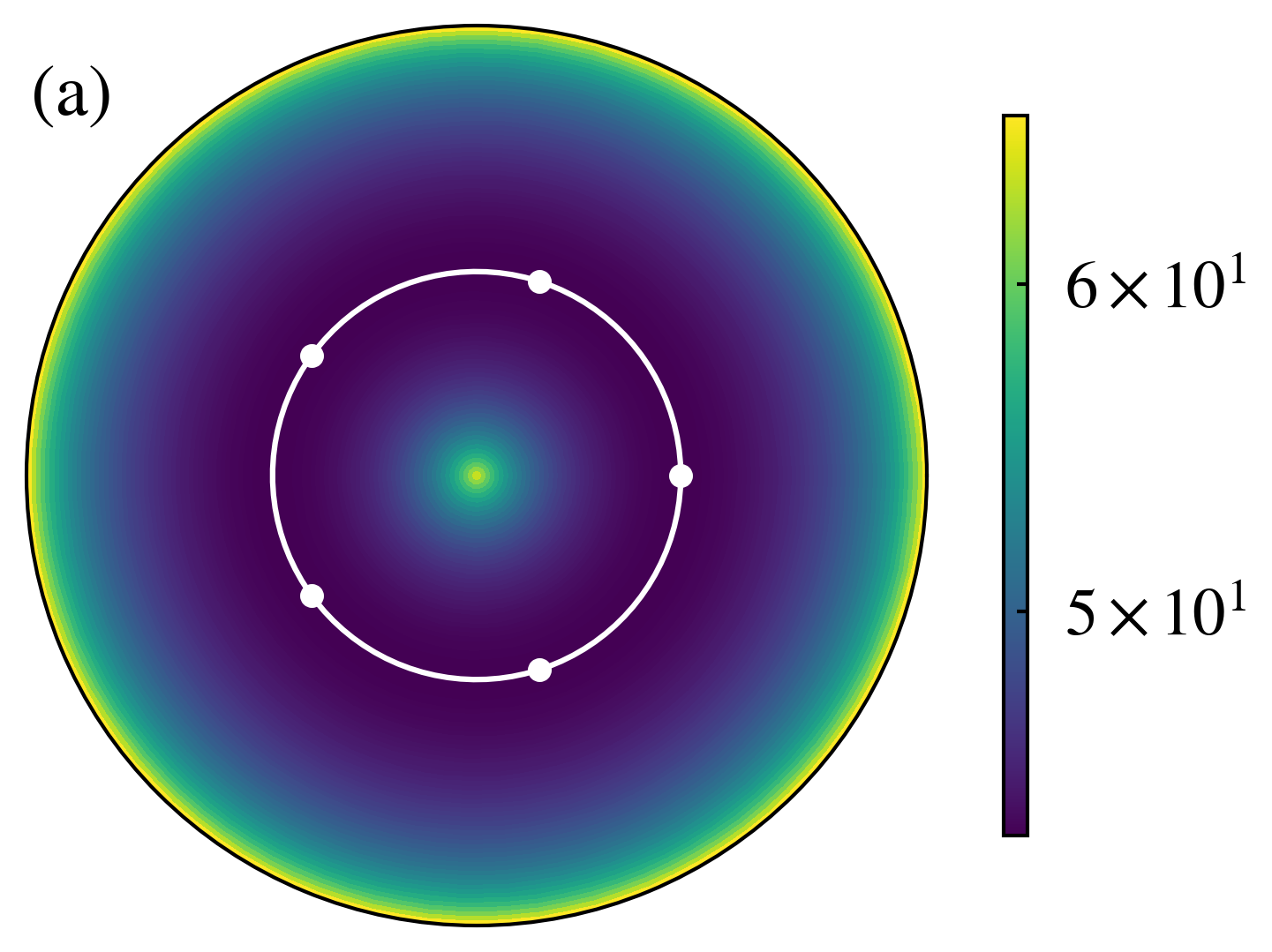}
    \includegraphics[width=\columnwidth]{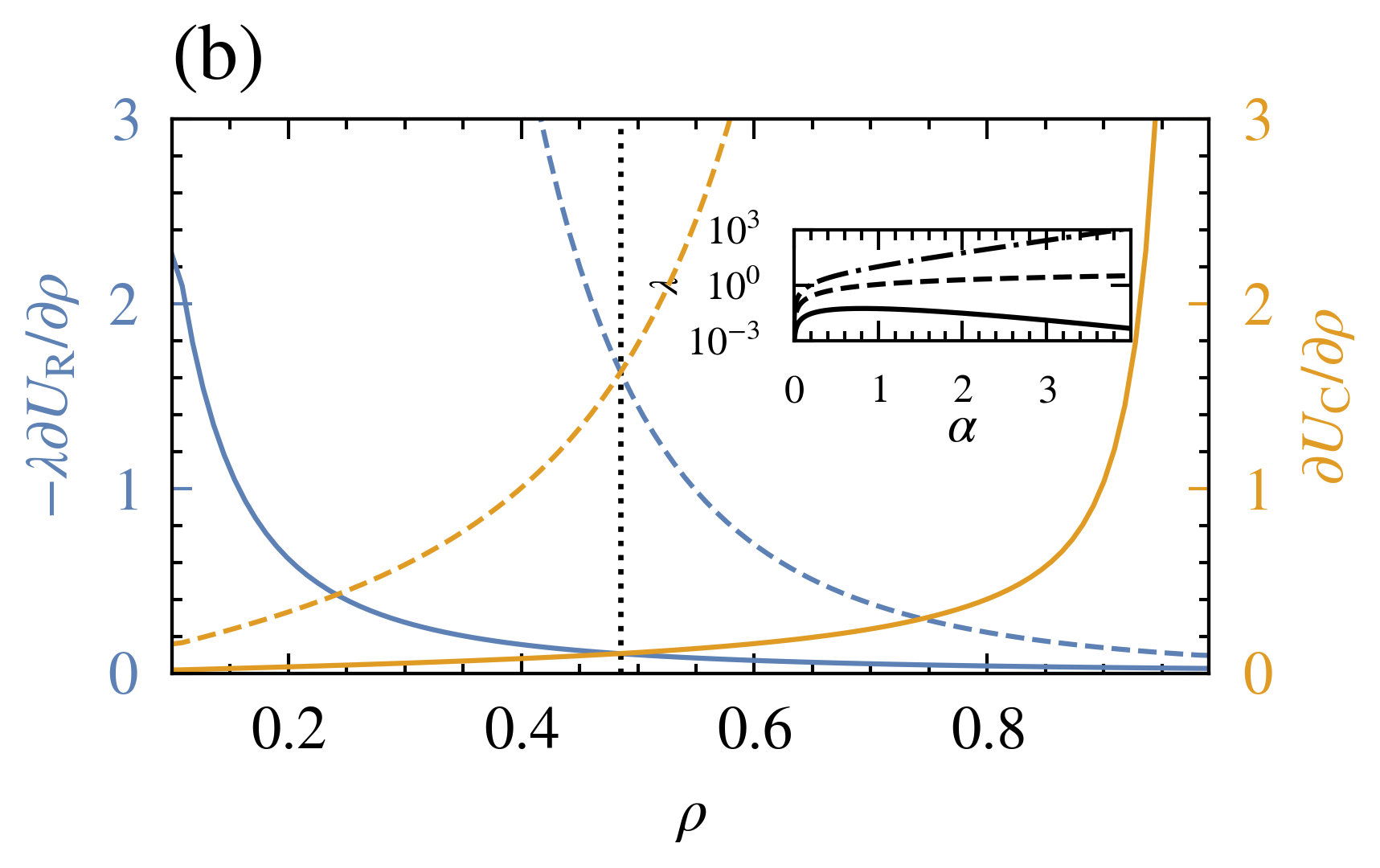}
    \includegraphics[width=\columnwidth]{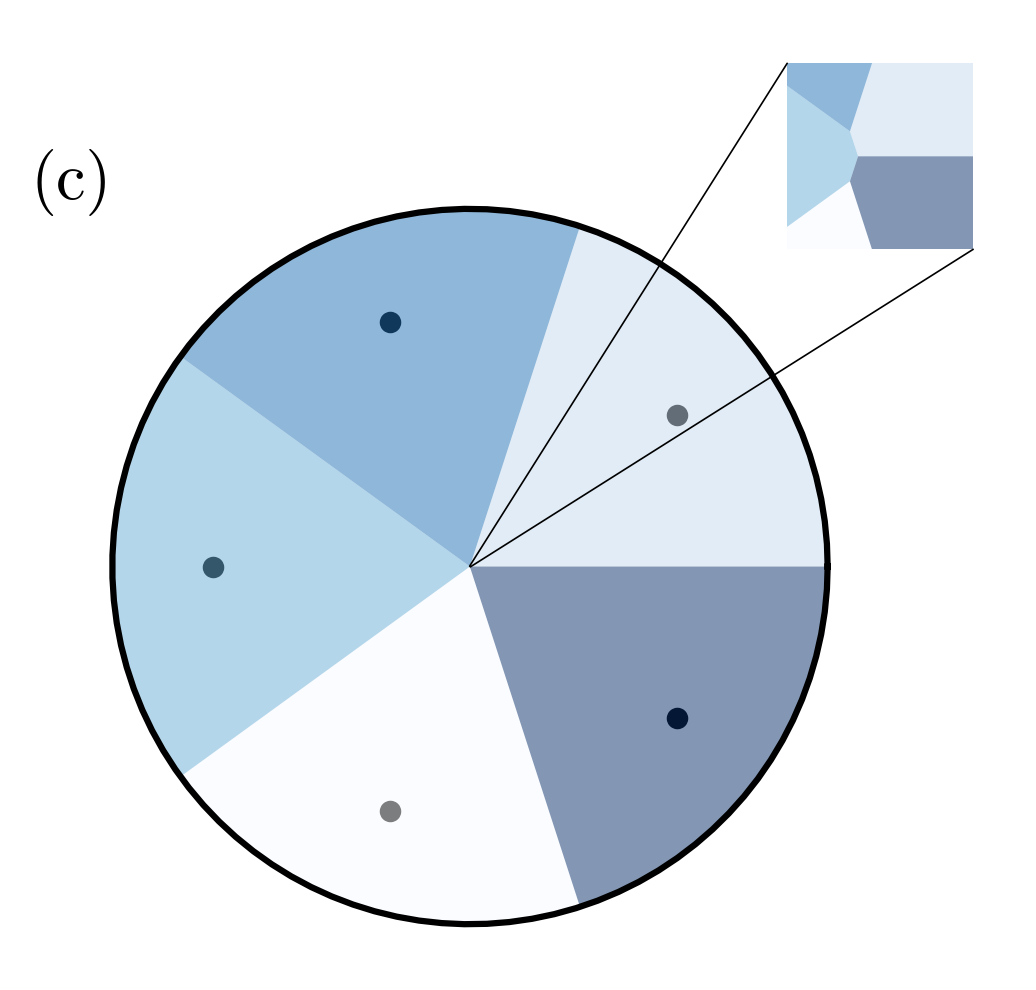}
    \includegraphics[width=\columnwidth]{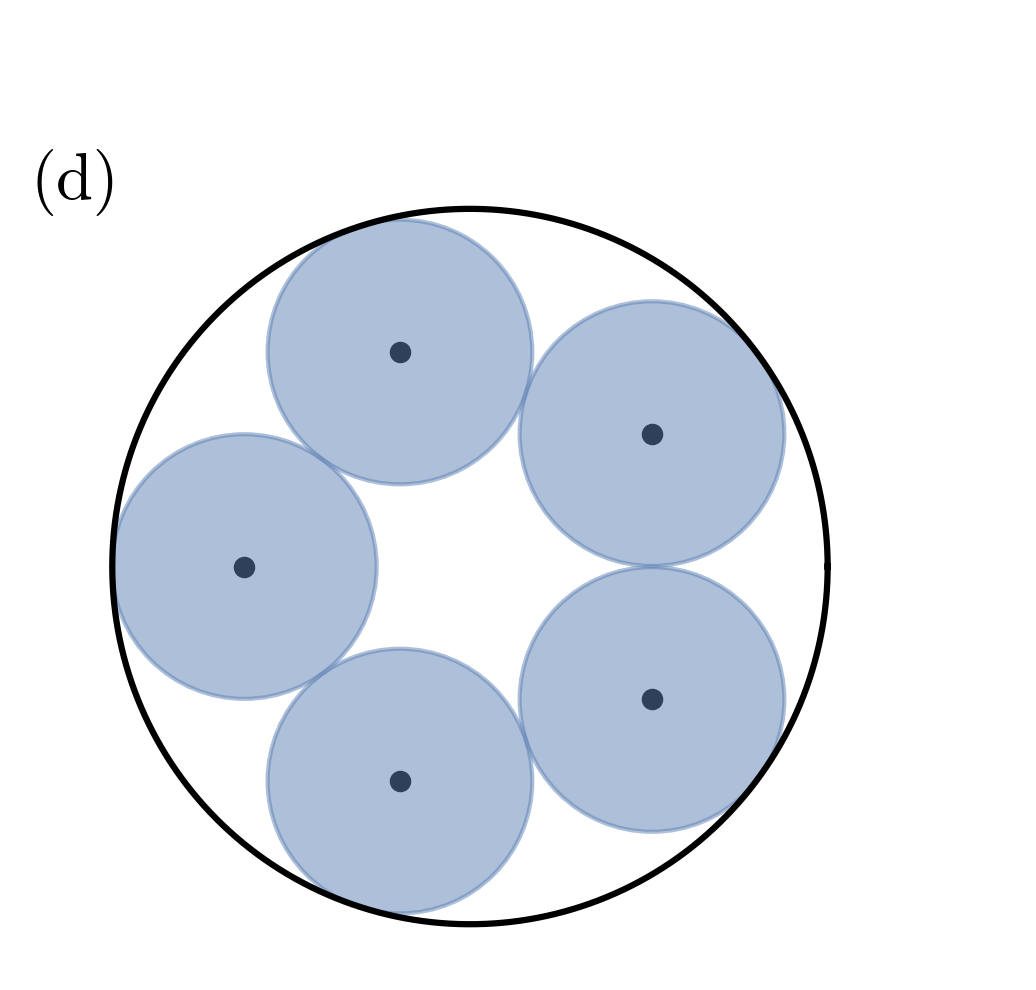}    
    \caption{(a) Colour plot indicating the total potential energy $U_{\mathrm{tot}}$ given by Eq. \eqref{eq:Utot}  for $N=5$ particles (seen as white dots) around a circular ring of radius $\rho$ and by a total of $M=36$ particles (not shown) homogeneously distributed at the circular boundary of unit radius. The value of $\lambda=1$ was chosen arbitrarily and $U_{\mathrm{R}}$ and $U_{\mathrm{C}}$ are defined by Eqs. \eqref{UR} and \eqref{UC} with $\alpha=1$. The white ring on which the $N$ particles are located indicates the equipotential. The colour code is such that bright (dark) colours indicate relatively high (low) potential values. (b) Radial plots (in arbitrary units) of the derivatives $-\lambda \, {\partial U_{\mathrm{R}}/\partial \rho}$, shown in blue (on the left axis), and ${\partial U_{\mathrm{C}}/\partial \rho}$, shown in orange (on the right axis). The solid (dashed) lines are for $\alpha=1$ and $\lambda=1$ ($\alpha=3$ and $\lambda=2.2$). The vertical dotted line indicates that the solid lines cross at exactly the same value of $\rho=0.49$ as the dashed lines. The inset displays the relationship between $\lambda$ and $\alpha$ required to have the $N=5$ particles at equilibrium at values $\rho=0.2$ (solid line),  $\rho=0.5$ (dashed line); $\rho=0.8$ (dot-dashed line). (c) Voronoi structure obtained through a simple energy minimisation of the guiding model with $N=5$ and $\lambda$ selected through the criterion of minimizing the variance in the areas of the Voronoi cells. The central part, enlarged in the inset, is topologically equivalent to the corresponding soap bubble experiments in Table \ref{t:comparison}. (d) Hard-sphere equilibrium configuration obtained through the same energy minimisation procedure, but this time by adjusting $\lambda$ to maximise the volume fraction of the cluster.} 
\label{fig:schematic-diagram}
\end{figure*}

These $M$ particles are uniformly distributed around the perimeter of a circle of radius $R$ and fixed, whereas the $N$ cluster-composing particles are allowed to move toward their equilibrium positions. Fig.~\ref{fig:schematic-diagram}(a) displays a schematic representation of this cluster with $N=5$ (white dots)  and $M=36$ (not shown). Thanks to the circular symmetry combined with the central nature of the interactions, the cluster particles lie on a ring of radius $\rho$ around the centre. The value of $\lambda$ determines the radius $\rho$, which gets larger (smaller) as $\lambda$ increases (decreases). For the case illustrated in Fig. \ref{fig:schematic-diagram}(a), finding $\rho$ consists of carrying out an energy balance between the repulsive and the confining contributions to the total potential energy. More specifically, the total potential energy of a particle is 
\begin{equation}
    U_{\mathrm{tot}}(\rho) = U_{\mathrm{C}}(\rho) + \lambda U_{\mathrm{R}}(\rho)\,,
    \label{eq:Utot}
\end{equation}
where 
\begin{equation}
U_{\mathrm{R}}(\rho)= \frac{1}{2^{\alpha/2}\rho^\alpha}\sum_{j\neq1}^{N} \left[1-\cos\left(\frac{2\pi j}{N}\right)\right]^{-\alpha/2}\,,
\label{UR}
\end{equation}
and
\begin{equation}
U_{C}(\rho)=\sum_{k=1}^{M} \left[1 + \rho^2 - 2\rho\cos\left(\frac{2\pi k}{M}\right)\right]^{-\alpha/2}\,.
\label{UC}
\end{equation} 

Eq. \eqref{UR} corresponds to the repulsion felt by one particle due to all the others in the cluster whereas Eq. \eqref{UC} comes from the confining forces exerted by the $M$ particles on the boundary. The indices $j$ and $k$ refer to the cluster and boundary particles, respectively. 
Being in equilibrium means that $\partial U_{\mathrm{C}}/\partial\rho = -\lambda \partial U_{\mathrm{R}}/\partial\rho$.
%\begin{equation}
%\frac{\partial U_{C}}{\partial\rho} = -\lambda\frac{\partial U_{R}}{\partial\rho}\,.
%\label{equilibrium}
%\end{equation}

It is worth mentioning that, at this point, no restrictions are being made on whether this equilibrium configuration falls into a local or a global minimum of the total potential energy $U_{\mathrm{tot}}$. Furthermore, despite using $\alpha=1$ for generating Fig.~\ref{fig:schematic-diagram}(a), the experimental results appearing on the leftmost column of Tab. \ref{t:comparison} are all very similar, even though none of them are mediated by the Coulomb interaction.

Regarding the generic forms of $U_{\mathrm{R}}$ and $U_{\mathrm{C}}$ written in terms of $\alpha$, we impose no constraint on this exponent other than it having to be positive. %Of course the Coulomb interaction is recaptured in the case of $\alpha=1$. Despite the apparent complexity of the expression above, it is possible to justify that the equilibrium condition will arise as a result of the competition between the two terms on the right-hand side. In fact, 
It is then straightforward to see that the derivative of the repulsive term diverges in the limit that $\rho \rightarrow 0$, decaying monotonically to a finite value when $\rho \rightarrow 1$. On the other hand, the derivative of the confining term vanishes at $\rho=0$ and increases as $\rho \rightarrow 1$. Therefore, there will always be a value of $0\leq\rho\leq1$ for which $\partial U_{\mathrm{tot}}/\partial\rho = 0$. This is illustrated in Fig. \ref{fig:schematic-diagram}(b) , which plots $-\lambda\partial U_{\mathrm{R}}/\partial\rho$ (solid blue line) and $\partial U_{\mathrm{C}}/\partial\rho$ (solid orange line) for the Coulomb case. Assuming $\lambda=1$ in this instance, both curves cross at $\rho=0.49$. A vertical dotted line indicates where the crossing occurs.

Let us now consider $\alpha
\neq 1$. %Because $\rho<R$,  the smaller the exponent $\alpha$ appearing in Eqs. (\ref{eq:pairwise}), (\ref{UR}) and (\ref{UC}), the more short-ranged the interaction becomes. 
Since the repulsive contribution is modulated linearly by the parameter $\lambda$, it is always possible to find a suitable value for that quantity to keep the equilibrium position unchanged. In fact, the dashed lines of Fig. \ref{fig:schematic-diagram}(b) correspond to  $\alpha=3$, where $\lambda=2.2$ has been adjusted to maintain the solution at $\rho=0.49$, {\it i.e.}, such that both curves cross right where the vertical dotted line is placed. The inset of Fig. \ref{fig:schematic-diagram}(b) displays the relationship between $\lambda$ and $\alpha$ required to sustain the equilibrium position unchanged for three separate values of $\rho$, namely $\rho=0.2$ (solid line), $\rho=0.5$ (dashed line) and $\rho=0.8$ (dot-dashed line). To summarise, we argue that a model with an adjustable repulsion-confinement ratio $\lambda$ can be employed to describe the exact equilibrium configuration of particle clusters in circular confinement, regardless of the underlying interaction. Strictly speaking, this argument works for a single shell but is sufficient to indicate the similarities in topology between different clusters.

The competition between repulsive and confining interactions is crucial for explaining particle cluster formation. As shown in Fig. \ref{fig:schematic-diagram}, the repulsive potential must decay from the cluster center, while the confining potential must increase radially, both monotonically. This raises the question of what happens when interactions become short-ranged ({\it e.g.}, contact forces) instead of following a power law. Two examples in Tab. \ref{t:comparison}, hard spheres and soap bubbles in confinement, demonstrate short-ranged interactions. For these, the absence of forces when objects are not in contact requires a modified strategy to determine the appropriate value for the parameter $\lambda$, different from the method in Fig. \ref{fig:schematic-diagram}(b). We will discuss this next.

%as follows: (i) for the hard spheres we simply impose that $\lambda$ must generate the maximum volume fraction (or the  minimum porosity); (ii) for the soap bubbles, we impose that the areas of the individual bubbles must be the same. Remarkably, these criteria enable us to generate equilibrium configurations that match very well with most of those found experimentally, both for the close-packing of hard spheres and for bubbles. 
 
For the hard spheres, for example, it is possible to find close-packing solutions by following a simple procedure. We minimise the total energy defined in Eq. (\ref{eq:Utot}), with any value of $\alpha$, say $\alpha=1$. We then proceed to find the value of $\lambda$ by imposing as an additional criterion that the packing fraction (area of $N=5$ circles divided by confinement area) must be maximum. Fig. \ref{fig:schematic-diagram}(d) shows the solutions obtained for the case of $N=5$.  This corresponds to the packing fraction $\phi=0.6845$, which coincides with the exact value reported in Ref. \cite{packomania}. 

A similar procedure is employed for the case of bubbles, the main difference being the criterion imposed to specify the parameter $\lambda$.  It is convenient to draw the particles as points (obtained through energy minimisation with an arbitrary exponent $\alpha$) together with the Voronoi diagram, that is, a representation in which the space is divided into polygonal-shaped regions that are closest to each of the particles. Besides being a valuable visual aid, the Voronoi cells reflect the number of nearest neighbours a given particle has (through the number of edges), as well as the particle density (through the reciprocal of the polygonal area). 

Our bubble experiments (Tab. \ref{t:comparison} and Sec. \ref{s:experiments}) were conducted with monodisperse bubbles. In our model, we can impose equal Voronoi cell areas $A_i^{\mathrm{vor}}$ by finding a value of $\lambda$ which minimises the area dispersion $\sigma^2 = (1/N)\sum_{i=1}^N(A_i^{\mathrm{vor}} - \langle A^{\mathrm{vor}}\rangle)^2$.
This is shown in Fig.~\ref{fig:schematic-diagram}(c) for the case $N=5$. It is important to stress that this simplified use of the guiding model serves in no way as a replacement to the standard way of describing the geometry of liquid foams, described in Sec. \ref{ss:simulation-foams}. The picture described here in terms of polygonal cells disregards the pressure differences between neighbouring bubbles, which lead to cell walls being arcs of circles and yet, remarkably, is capable of obtaining the same topologies, by which we mean the number of neighbours of each region, as the results depicted in Table \ref{t:comparison}.  The inset in Figure \ref{fig:schematic-diagram}(c) shows that what at first glance may seem like a distinct topology in this highly-symmetrical particle arrangement, is actually identical to the one seen in Table \ref{t:comparison}. Therefore, it is yet another piece of evidence supporting our claim that seemingly disparate physical systems may display very similar arrangements. 

%As seen in the inset of Fig.\ref{fig:schematic-diagram}(b), different values of $\alpha$ may display exactly the same spatial arrangement for their particles if the parameter $\lambda$ is adjusted accordingly. Therefore, this suggests that even in the case of short-range potentials like the ones associated with the hard-spheres and the soap bubbles, for example, the correct structure can be recovered even if the power-law potential of Eq.(\ref{eq:pairwise}) is reinstated. This enables one to use values for $\alpha$ and $\lambda$ that reduce the computational inaccuracies. Fig.\ref{fig:lambda}(a) illustrates this for the impenetrable spheres where curves for the volume fraction $\Phi$ as a function of $\lambda$ are shown for two different values of $\alpha$ as well as for the short-range potential described by Eqs.(\ref{eq:usr}) and (\ref{eq:uc-sr}). Each one of the three curves reaches the same maximum value of $\Phi_{max}=7/9$ but at distinct values of $\lambda$. Any of the three gives rise to the same close-packed diagram shown in Fig.\ref{fig:lambda}(b). Likewise, the same occurs for the bubbles. Fig.\ref{fig:lambda}(c) plots the standard deviation $\sigma$ of the areas of the individual Voronoi cells as a function of the ratio $\lambda$. It contains three different curves, each one of them with similar minima $\sigma_{min}$ at different values of $\lambda$. All these minima generate the same  structure shown in Fig. \ref{fig:lambda}(d).

\begin{figure}
    \centering
    \includegraphics[width=\columnwidth]{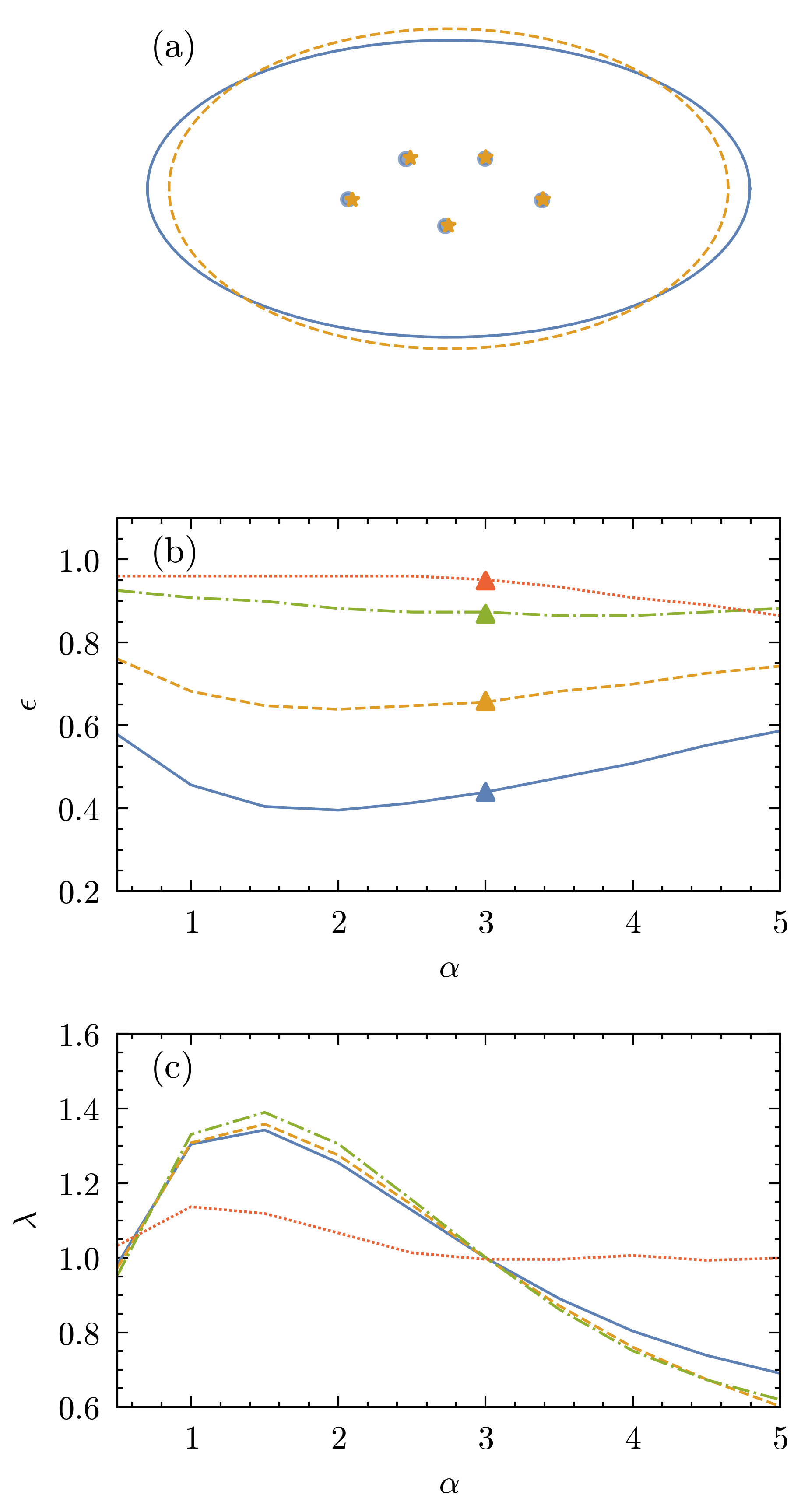}
    \caption{(a) Example of identical equilibrium configuration for $N=5$ particles under different interaction potentials.
    The orange stars correspond to the  Coulomb case ($\alpha=1$, $\lambda=1.3$, $\epsilon=0.91$) and the blue circles to the magnetic case ($\alpha=3$, $\lambda=1$, $\epsilon=0.87$), both with $M=36$ and ellipses in the same colour.
    (b) This plot indicates the relationship between the eccentricity of the confining ellipse and the exponent $\alpha$ required to maintain a given equilibrium configuration. The four points highlighted at $\alpha=3$ are the elliptical eccentricity values seen in Table \ref{t:comparison}, namely $\epsilon = 0.44$ (blue triangle), \, $0.66$ (orange triangle), \, $0.87$ (green triangle) and $0.95$ (red triangle), for the case of $N=5$ floating magnets. (c) This plot indicates how the parameter $\lambda$ must change with $\alpha$ to maintain the same four equilibrium configurations highlighted in panel (b). }
    \label{fig:elliptical-cases}
\end{figure}

%\begin{figure}
%    \centering
%    \begin{subfigure}[b]{0.48\textwidth}
%    \includegraphics{figures/intro_figures/ellipse_fig_2a.png}
%    \end{subfigure}
%    \begin{subfigure}[b]{0.48\textwidth}
%    \includegraphics{figures/intro_figures/ellipse_fig_2b.png}
%    \end{subfigure}
%    \caption{
%\end{figure}

%\begin{figure} [hbt!]
%    \centering    \includegraphics[width=0.8\linewidth]{figures/mag_simulations/WhatsApp Image 2025-04-02 at 10.07.23.jpeg}
%    \includegraphics[width=0.5
%    \linewidth]{figures/mag_simulations/WhatsApp Image 2025-04-02 at 10.24.37.jpeg}
%    \includegraphics[width=1.0\linewidth]{figures/mag_simulations/Screenshot 2025-04-02 at 13.25.07.png}
%    \includegraphics[width=0.5
%    \linewidth]{figures/mag_simulations/Screenshot 2025-04-02 at 13.24.15}
%    \caption{(a) Volume fraction $\Phi$ as a function of $\lambda$ for the cases $\alpha=10^{-2}$,  $\alpha=1$ and $\alpha=3$. The maximum of $\Phi$ is found at $7/9$ for all three cases. Panel (b) displays the close-packed configuration generated by any one of the three maxima seen in panel (a). Panel (c) plots the standard deviation of the areas of the individual Voronoi cells as a function of the ratio $\lambda$ for the same $\alpha$ values used to generate the curves of panel (a), all leading to the same minimum. Panel (d) draws the obtained equilibrium positions for the particles together with the corresponding Voronoi diagram that mimics very closely the bubble structures. Each of the minima of panel (c) gives rise to the same diagram. }
%    \label{fig:lambda}
%\end{figure}

\subsection{Elliptic confinement}

A similar argument to the one just presented applies when the symmetry of the confining field is not circular but elliptic. A single parameter $\lambda$ is not sufficient to determine the required equilibrium configuration. We use, in addition,  the eccentricity $\epsilon$ of the ellipse. Then, if the nature of the interaction changes, say from $\alpha=1$ (Coulomb) to $\alpha=3$ (magnetic dipolar), it is possible to maintain the same equilibrium configuration of the particles if the eccentricity of the confining ellipse is also changed. This is shown in  Fig.~\ref{fig:elliptical-cases}(a), where two sets of $N=5$ particles interacting via a different potential are confined within ellipses of different eccentricities. When superposed, the equilibrium positions of the two sets are the same, bar small numerical deviations. In other words, similarly to the circular case seen in Fig.~\ref{fig:schematic-diagram}, it is possible to identify the same equilibrium configurations of particle clusters that are governed by distinct interactions. In the same spirit, the thick solid line in  Fig.~\ref{fig:elliptical-cases}(b) indicates what eccentricity values $\epsilon$ one would need to produce the same equilibrium configuration as in panel (a) for values of $\alpha$ other than $1$ and $3$. In addition, three other curves are shown reflecting the precise equilibrium configurations found for the floating magnets in Table \ref{t:comparison}. Finally, the four configurations seen in panel (b) will remain the same if the parameter $\lambda$ changes with  $\alpha$ as shown in panel (c). This feature once again highlights a high degree of universality in the formation of self-assembled clusters, suggesting that these similarities are likely to occur even beyond the circular and elliptical confinements considered here. 

In summary, the guiding model encapsulates mutually repelling particles under the action of confining forces. Once the confinement-repulsion ratio is adjusted together with the eccentricity, it is possible to find the same equilibrium configurations regardless of the underlying interaction potentials. In the next section we describe how we were able to experimentally generate similar equilibrium structures in clusters governed by very different physical interactions (using hard spheres, magnetic particles, and bubbles; for Coulomb interactions see \cite{SaintJeanEtal2001,SaintJeanGuthmann2002}), all of which under the action of circular and elliptical confining fields.

\section{Experimental realisations of packings in ellipsoidal confinement}
\label{s:experiments}

In both the experiments described in this section and the simulations of section \ref{s:simulations}, we restrict ourselves to elliptical confinements of five different values of eccentricity, $\epsilon \in \{0,0.44, 0.66, 0.87, 0.95\}$, where $\epsilon = \sqrt{1 - (b/a)^2}$, with $a$ and $b$ denoting respectively the lengths of the semi-major and semi-minor ellipse axes. Experimental results are described for $N=5$ and $N=10$ particles.

\subsection{Experiments with hard spheres}
\label{ss:hard-spheres}
Experimenting with dense packings of hard spheres is probably the simplest way to illustrate the role that confinement plays in the formation of ordered arrangements. The structures encountered for given values of eccentricity $\epsilon$, and number of spheres $N$, will also be identified in our experiments with floating magnets and bubbles.

Since we are only considering a monolayer of spheres, the packings are equivalent to the problem of the densest two-dimensional packings of non-overlapping hard disks in confinement. In this context, the problem may be stated as follows. Given an ellipse of area $A$ and eccentricity $\epsilon$, what is the area $A_c$, of identical circles that maximise the packing fraction $\phi = NA_c/A$, for a given number $N$ of circles.

For the case of a {\em circular} confinement, i.e. $\epsilon=0$,  there is an online database~\cite{packomania}  of conjectured optimal packings, obtained computationally, for values of $N$ up to 2604. However, systematic computational results for confining ellipses appear to be limited, due to the much more demanding computation in determining circle-ellipse overlaps. Birgin {\em et al.} describe a nonlinear optimization procedure which computes dense packings of more than 300 circles within an ellipse \cite{BirginEtal2013}. However, their main interest relates to algorithm efficiency.
Amore {\em et al.} compute the variation in packing density with the length of the semi-major axis for three values of $N$ (37, 61, 91) \cite{AmoreEtal2023}; some further simulation results are displayed  by \cite{Friedman}.

Here we describe an illustrative experimental approach to obtain dense sphere packings in confinement.
It involves the use of sets of metal spheres (with a different sphere diameter in each set) and 3D-printed elliptical frames of different areas for the five different values of eccentricity, as stated above. We validate our results by comparing with conjectured optimizers for $N=5, \epsilon = 0.867$ and $N=10, \epsilon = 0.6539$~\cite{Friedman}.

To obtain a dense sphere packing for a given elliptical frame, we proceed as follows. The frame is placed on a horizontal surface, and we manually dispose spheres of a given size into it.
This process does generally not result in a dense packing, i.e. there comes a point when there is enough room available for the spheres to rattle, but not enough room for the addition of a further sphere, without losing contact with the horizontal surface. We thus start the process again, but now with slightly larger, or smaller spheres. By trial and error we minimise the amount of free space to arrive at a (near) dense configuration. 

Table \ref{t:comparison}
shows photographs of the arrangements that we consider sufficiently space-filling to demonstrate a dense packing. For the circular confinement, $\epsilon=0$ (for which we have simulated data to compare with \cite{packomania}),  we reproduce the conjectured densest arrangements for both $N=5$ and $N=10$. Our experimentally obtained values for the packing fractions are $\phi_5 = 0.68 \pm 0.01 $ and  $\phi_{10} = 0.69 \pm 0.01 $; this is consistent  with the published results $\phi_5 = 5 / (1+1/\sin(\pi/5))^2 \simeq  0.685 $ and  $\phi_{10} \simeq 0.688$ \cite{packomania}. Also our results for $N=5, \epsilon = 0.87, \phi=0.68\pm0.01$ and $N=10, \epsilon = 0.66, \phi=0.77\pm0.01$ are consistent with those of simulations ($\phi_5 = 0.6869, \phi_{10} = 0.7797$) \cite{Friedman}.

%\remSC{In the comparison section you discuss contacts; how do you record contacts in the experiments?} \remSH{visual observation}

%\begin{table*}[hbt!]
%    \centering
%    \begin{tabular}{|c|c|c|c|c|c|}
%\hline
%$\epsilon$      & 0 & 0.44 & 0.66 & 0.87 & 0.95 \\
%\hline
%
%N=5 & \includegraphics[width=0.15\textwidth, angle = 0]{figures/Hard sphere data/HS files/e=0/N=5.jpeg} & &\includegraphics[width=0.15\textwidth, angle = 0]{figures/Hard sphere data/HS files/e = 0.66/N=5.jpeg}  & &\includegraphics[width=0.1\textwidth, angle = 90]{figures/Hard sphere data/HS files/e =0.95/N=5.png} \\
%\hline
%
%N=10     & & \includegraphics[width=0.15\textwidth, angle = 0]{figures/Hard sphere data/HS files/e=0.44/N=10.png} & \includegraphics[width=0.15\textwidth, angle = 0]{figures/Hard sphere data/HS files/e = 0.66/N=10.png}& \includegraphics[width=0.1\textwidth, angle = 90]{figures/Hard sphere data/HS files/e = 0.87/N=10.png}& \includegraphics[width=0.1\textwidth, angle = 90]{figures/Hard sphere data/HS files/e =0.95/N=10.png} \\
%\hline
%
%N=20     & & & & \includegraphics[width=0.18\textwidth, angle = 0]{figures/Hard sphere data/HS files/e = 0.87/N=20.png} & \includegraphics[width=0.22\textwidth, angle = 0]{figures/Hard sphere data/HS files/e =0.95/N=20.png} \\
%\hline
%
%\hline
%
%    \end{tabular}
%    \caption{Dense arrangements of hard spheres in elliptical confinement for five different values of eccentricity, $\epsilon$. The structures found for $\epsilon=0$ (circular confinement) match those of the simulations \cite{packomania}.}
%    \label{t:exp-hard-spheres}
%\end{table*}

\subsection{Experiments with floating magnets}
\label{ss:exp-floaters}

\begin{figure*}
(a)\includegraphics[width=0.97\columnwidth]{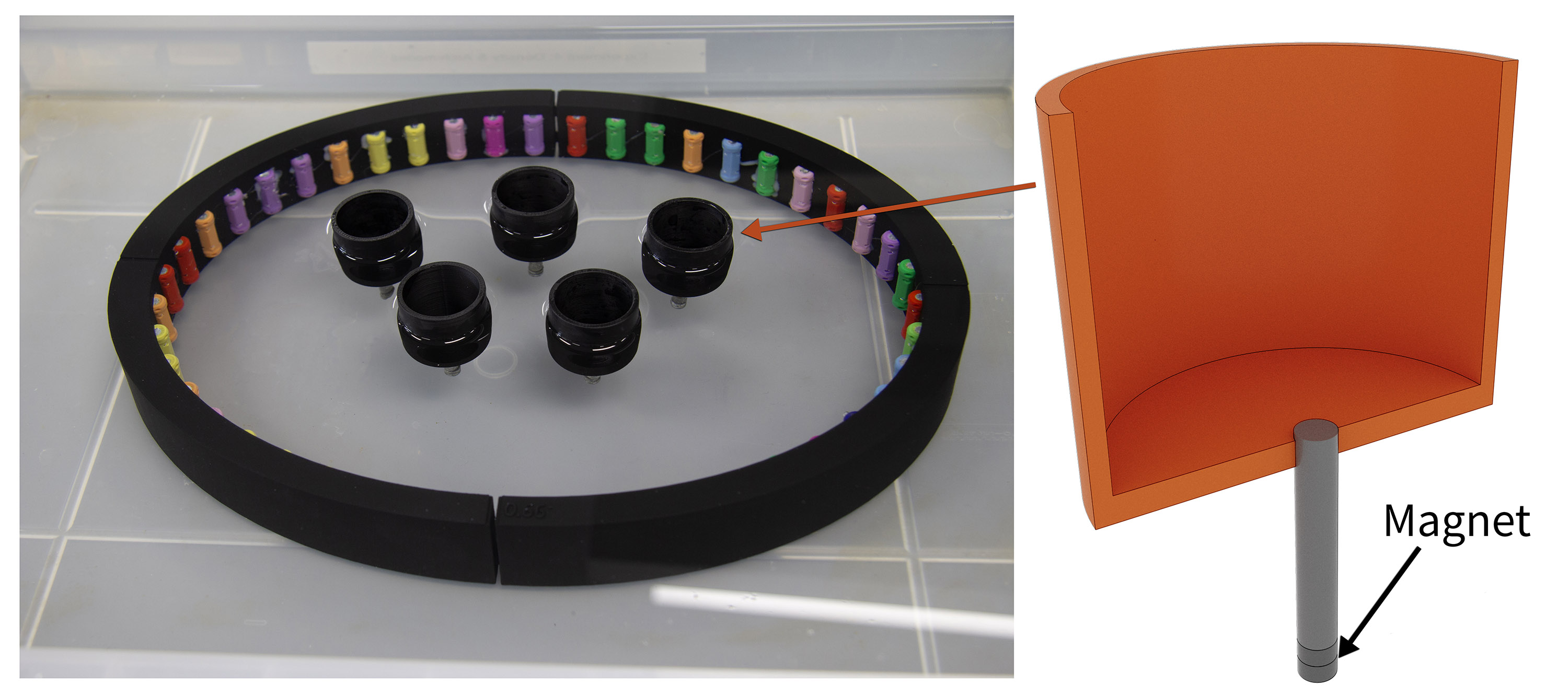}
(b)\includegraphics[width=0.97\columnwidth]{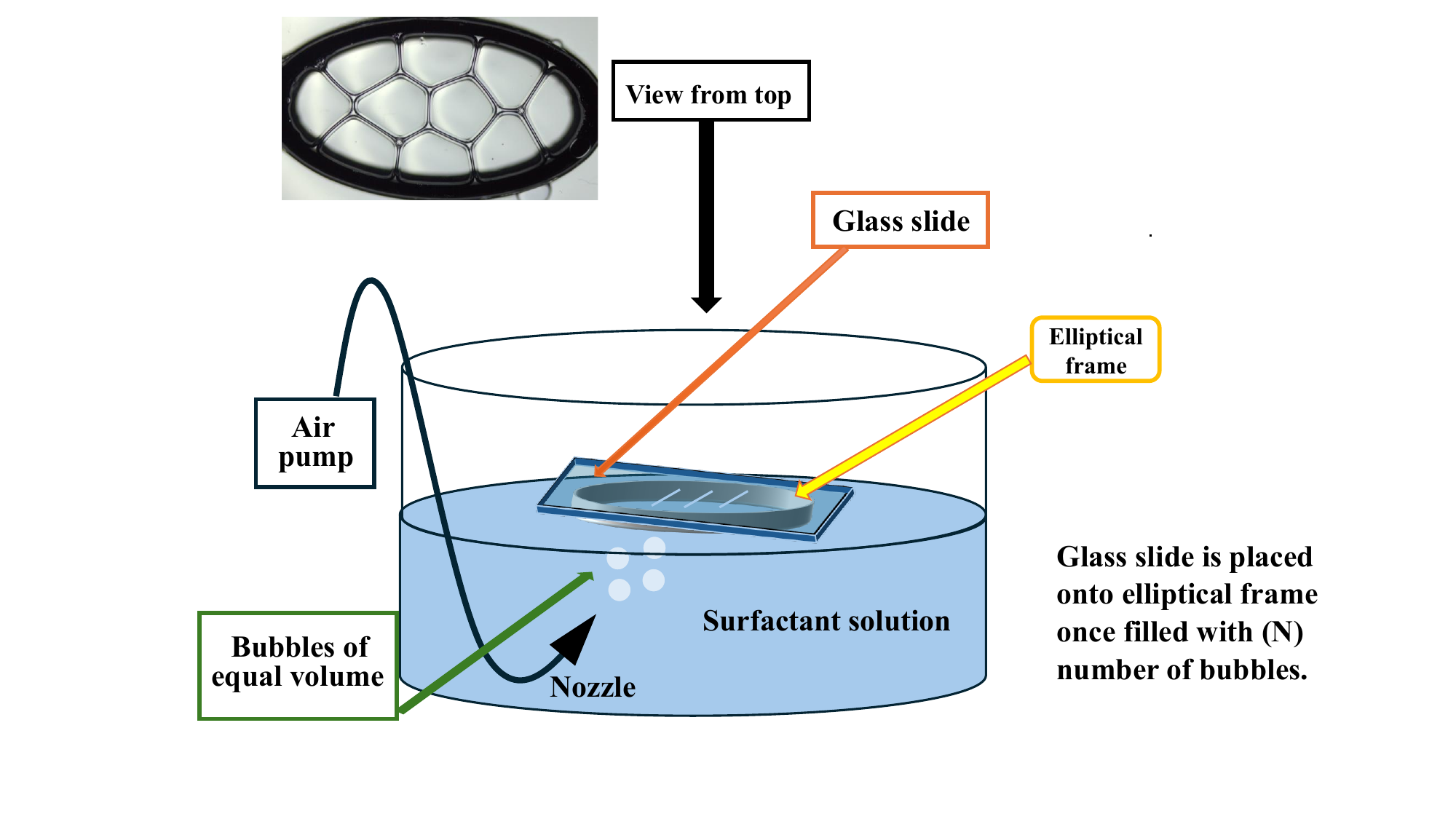}
\caption{\label{fig:foam_setup} Experimental set-ups for the creation of elliptically confined clusters of magnetic floaters, or bubbles.  (a) A cluster of five magnetically repelling floaters in water, confined by a ring of elliptically arranged magnets.  The view of (the cross-section of) one of the floaters shows the small disc magnets that are attached at the bottom of the metal ``keel''. (b) Monodispersity of the bubbles is ensured by having a constant air flow. Once the elliptical frame is entirely filled with bubbles, a glass cover is placed over it, contacting the bubbles. This results in a two-dimensional cellular structure, when viewed from above.}
\end{figure*}

Experiments with mutually repelling magnets, floating on water, and subject to an external magnetic confining field, are generally traced back to the 19th century physicist Alfred M. Mayer \cite{Mayer1878a, Mayer1878b}. His main interest, and that of many others who followed, e.g. \cite{Derr1909,RiverosEtal2004, NemoianuEtal2022, deLimaEtal2024}, was the identification of equilibrium structures for a {\em circularly} symmetric confining field.
Here we report experiments in which the confining field has elliptical symmetry, with the same values of eccentricity as given above.

Our experiments were performed using floaters made from 3D-printed hollow cylinders, open at the top (diameter 22.4 mm, height 25 mm), as illustrated in Fig. \ref{fig:foam_setup}(a). A metallic rod (length 23 mm, diameter 4.5 mm) was led through a hole at the bottom, and two small neodymium disc magnets (diameter 5 mm, each of thickness 1.4 mm) (purchased from \cite{magnet}) were attached to it at the bottom end. 
When placed into a pool of water, the magnet-fitted cylinders float, with the magnets being submerged in the water.

We also 3D-printed four elliptical frames, with the same cross-sectional area of $\pi$ cm$^2$ but different eccentricities. 
Bar magnets (`toy magnetic building sticks') were attached along the inner perimeter of each frame, in opposite orientation to that of the floating magnets, and thus providing a confining magnetic field. The same number of magnets, 42, was used for all of the frames; since the length of the perimeter of an ellipse increases with eccentricity, $\epsilon$, this results in a number density of magnets which decreases with $\epsilon$.

The magnet-fitted elliptical frame rests on the bottom of a pool of water, and its rim projects above the water level. Magnetic floaters were then placed one by one into the frame and allowed to settle into equilibrium positions. After about one minute, photographs of the resulting configurations were taken from above.

Results for $N=5$ and $N=10$ are shown in Tables \ref{t:comparison} and \ref{t:exp-bubbles-alternatives}.
In contrast to our hard sphere packings, the floating magnets are in contact with neither each other nor the confining frame. Nevertheless, the arrangements of their centres are similar to those of the hard spheres. We will further discuss similarities and differences in Section \ref{s:comparison}.

\subsection{Experiments with monolayers of bubbles}
\label{ss:foam-expts}

Arrays of soap bubbles floating on a liquid (introduced by Bragg and Nye \cite{BraggNye1947} to model crystal structure) serve as another simple quasi-2D experimental system to demonstrate the formation of structures with a now familiar topology when in confinement. Similar to the hard spheres of Section \ref{ss:hard-spheres}, bubbles experience repelling forces only when they are in contact with each other or with the confining frame.

In the following, we describe experiments performed with monolayers of equal-volume bubbles, confined within 3D-printed elliptical frames of equal area but different eccentricity. The bubbles were produced by blowing air at constant pressure through a nozzle into a pool of aqueous surfactant solution; see Fig.~\ref{fig:foam_setup}(b). (We chose the commercial detergent ``Fairy Liquid'' as it is known to produce stable foams.) The elliptical frame is placed at the surface of the liquid pool, so that the rising bubbles aggregate within it, with their vertical extension slightly exceeding the height of the frame rim. Once the frame is filled with a monolayer of bubbles, a glass plate is used to cover it. Since the plate is also contacting the bubbles, this leads to a two-dimensional confined {\em covered} Bragg raft, as was pioneered in the experiments by \cite{VazFortes1997}). This raft is viewed from top, resulting in an image of a two-dimensional cellular structure. 

The number of bubbles, $N$, that can be placed within an ellipse of given area depends on the size of the monodisperse bubbles. However, the quasi-2D character of the experiment, i.e. the finite vertical dimension of the raft, allows for the accommodation of a range of bubble sizes for a given value of $N$.

We have gathered data for $N=5$ and 10, 
%and $20$ 
for a total of five different elliptical frames (eccentricity $\epsilon=0, 0.44, 0.66, 0.87, 0.95$) and fixed frame cross-section $A= \pi$ cm$^2$. The equivalent sphere diameters of the bubbles in use were 3.0mm ($N=5$), 2.8mm ($N=10$). 
%\remSH{probably not required, ... (N=20: 4.7mm bubble diameter, frame area $\Pi cm^2$) -- double check area for N=5,10 systems.} 
The height of the frame rim above the waterline was about 2.0mm.

For each pair of numbers of bubbles $N$ and eccentricity $\epsilon$, we conducted at least 10 experimental runs. This enabled the identification of several alternative bubble configurations. 
Table 
%\ref{t:exp-bubbles} 
\ref{t:comparison}
shows the subset of structures which correspond to configurations of minimal energy, as identified in our computer simulations of 2D foams, described in Section \ref{ss:simulation-foams}. Further configurations are shown in Appendix \ref{a:further-structures}, Table \ref{t:exp-bubbles-alternatives}.

\section{Computer simulations of Clusters in elliptical confinement}
\label{s:simulations}

\subsection{Simulations of two-dimensional foams}
\label{ss:simulation-foams}

\begin{figure}
(a)
\includegraphics[width=\columnwidth]{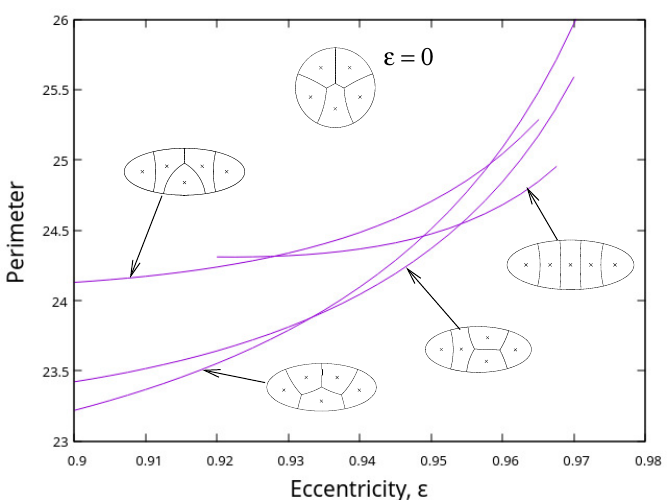}
(b)
\includegraphics[width=\columnwidth]{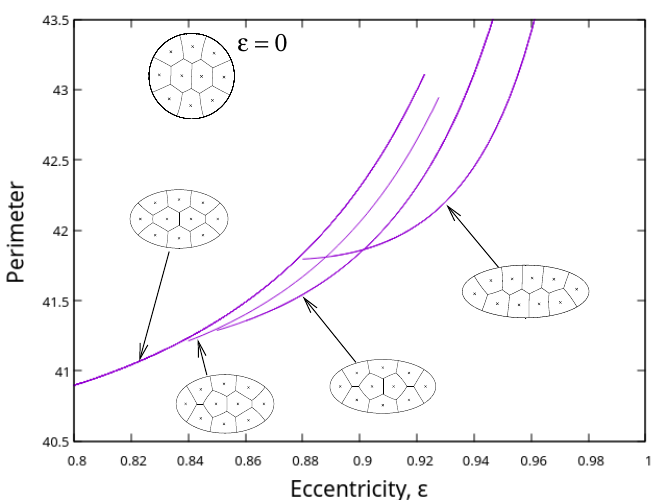}
\caption{\label{fig:foam_sim} 
Optimal arrangement of $N$ bubbles in an ellipse with varying eccentricity $\epsilon$. The value of the perimeter given includes the length of the elliptical frame. 
(a) $N=5$. The topology of the least-perimeter cluster is the same for $0 \le \epsilon < 0.93$, above which there are two further arrangements of the bubbles. (b) $N=10$. The topology of the least-perimeter cluster is the same for $0 \le \epsilon < 0.84$, above which there are several different arrangements of the bubbles, not all shown. 
}
\end{figure}

In an ideal two-dimensional dry foam, each bubble is enclosed by soap films consisting of circular arcs that
%. The soap films are smoothly curved -- by the Laplace Law they have the same pressure on each side, and hence constant curvature -- and 
meet in threes at $120^\circ$, and meet the confining walls at $90^\circ$~\cite{WeaireHutzler,mousse13}. We create monodisperse foam containing $N$ bubbles in an ellipse with eccentricity $\epsilon$ using the Surface Evolver~\cite{brakke92} in circular arc mode. 
%(Since the boundary of the ellipse is not well-represented by circular arcs, it is omitted and the missing length calculated directly.) \remSH{Simon, this is not very clear. Presumably, there is a film also around the frame, which is the one whose length you get straight from the perimeter of the ellipse. Could you look at the text above again and clarify this?}
The energy of a 2D foam is proportional to the total length of the soap films (the perimeter); we find local minima of the perimeter to high accuracy. We examine many local minima from which we conjecture the global minimum for each $N$ and $\epsilon$.

The method is based on the one described in~\cite{Coxf10}: for each $N$ and $\epsilon$, we find an approximation to the equilibrium position of $N$ particles confined in an ellipse and interacting via a Coulomb interaction potential (described in more detail in Section \ref{ss:simulation_floater}). These then act as seed points for a Voronoi construction, as described above. The Voronoi regions are then imported into Surface Evolver and treated as bubbles (with curved edges), and local minimum of the total perimeter is found by gradient descent.
%minimized. \sout{to find a local minimum}. \remSH{can this be rephrased, minimized to find a minimum?} 

This is followed by a Metropolis-like ``shuffling" procedure, in which we delete an edge,  either chosen at random or the edge with the shortest length, to see if the foam perimeter is decreased -- if it is, the new foam structure is kept, else it is rejected. This process results in a conjectured minimal foam for each ($N,\epsilon$). Finally, for each of these foams, we vary $\epsilon$ over a small range (roughly $\pm 0.2$) as a check that the foam structure of lowest perimeter has been found also at nearby eccentricities. These
%and record the perimeter over a range of values of $\epsilon$ , 
results are in Fig.~\ref{fig:foam_sim}.

%Method doesn't work well at large $\epsilon$ (greater than about $0.95$) because: (i) if I omit the ellipse walls I can't work out the content integral for the missing area when there is a single layer of bubbles, but (ii) if I include the ellipse walls they are very badly represented by circular arcs, so the accuracy is poor.

\subsection{Simulations of floating magnets}
\label{ss:simulation_floater}

To predict the arrangement of floating magnets, we treat them as classical particles that interact via a pair-wise potential and are subject to a confining potential, which is a particular case of the guiding model presented in Sec. \ref{energetics} with $\alpha = 3$. However, we stress that our method would be the same for the Coulomb case ($\alpha=1$)~\cite{BoltonRoessler1993,KongEtal2002,lailin99,bedanovpeeters94}.

\begin{figure*}
    \centering
    \includegraphics[width=\textwidth]{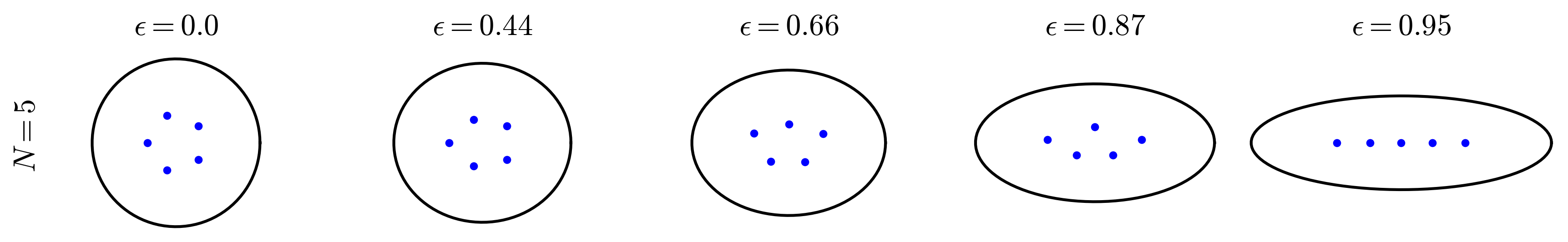}
    \includegraphics[width=\textwidth]{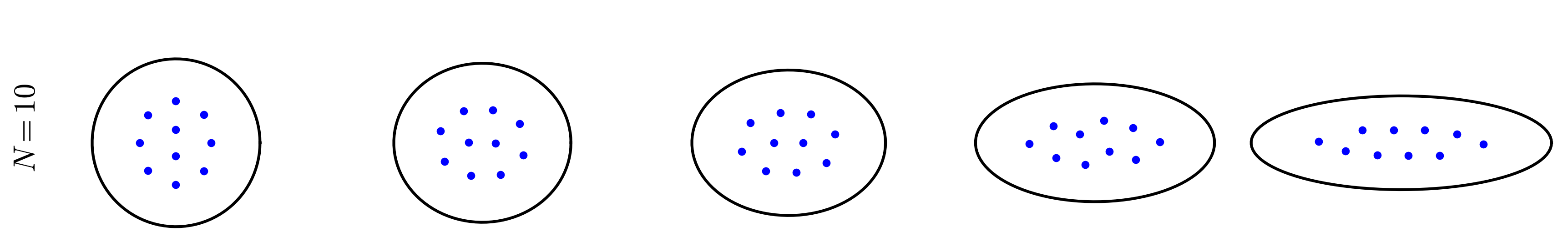}
    \caption{The equilibrium configurations for the floating magnets simulations for $N=5$ (top panel) and $N=10$ (bottom panel). The parameter $\lambda$ was calibrated from one experimental observation ($N=5, \epsilon=0$). The scale varies between figures with different eccentricities, but all figures share the same area.}
    \label{fig:magnet-simulations}
\end{figure*}

In this regard, we obtain equilibrium configurations of $N$ interacting magnetic particles possessing a magnetic moment $m$, which are geometrically confined in a two-dimensional domain by $M$ fixed magnets with magnet moment $m^{\mathrm{ext}}$ on the boundary of an ellipse. Assuming that all magnet moments involved are aligned with the $z$-direction, the total energy can be written as
\begin{equation}
	u(r) = \frac{\mu_0}{4\pi}\sum_{i=1}^N\sum_{k=1}^{M}\frac{mm^{\mathrm{ext}}}{r^{3}_{ik}} + \frac{\mu_0}{8\pi}\sum_{\substack{j = 1 \\ i\neq j}}^{N}\frac{m^2}{r^{3}_{ij}} \,, \label{eq:mag_energy}
\end{equation}
where $\mu_0$ is the magnetic permeability of the medium, and $r_{ik}$ is the distance between the $i$-th particle and $k$-th fixed magnet, while $r_{ij}$ correspond to the distance between two moving particles. For a fixed cross-section area $\pi ab$ and eccentricity $\epsilon$, as in  Sec. \ref{ss:exp-floaters}, the semi-axes are given by $a = (1 - \epsilon^2)^{-1/4}$ and $b = (1 - \epsilon^2)^{1/4}$.
This allows us to rewrite the total magnetic energy Eq. \eqref{eq:mag_energy} in a dimensionless form:
\begin{align}
	U&_{\mathrm{tot}} = \sum_{\substack{j = 1 \\ i\neq j}}^{N}\left[(\eta_i - \eta_j)^2 + (1+\epsilon^2)(\xi_i - \xi_j)^2\right]^{-3/2}\nonumber \\
    +& \frac{\lambda}{2}\sum_{i=1}^N \sum_{k=1}^{M} \left[(\eta_i - \eta_k)^2 + (1+\epsilon^2)(\xi_i - \xi_k)^2\right]^{-3/2}\,, \label{eq:floaters_total_energy}
\end{align}
where $U_{\mathrm{tot}} = 4\pi a^3 u/(\mu_0 mm^{\mathrm{ext}})$ with $\lambda = m^\mathrm{ext}/m$. Moreover, we have defined the dimensionless coordinates $\eta_i = x_i/a$ and $\xi_i = y_i/b$ which are constraints to $\eta_i^2 + \xi_i^2\leq 1$.

We use the generalised simulated annealing (GSA) method~\cite{gsa_tsallis_1996, gsa_tsallis_1996b, gsa_eff_2000, gsa_R_2013} to efficiently minimize Eq. \eqref{eq:floaters_total_energy} for the $2N$ equilibrium positions $(\eta_i, \xi_i)$. We proceed similarly as in Ref. \cite{deLimaEtal2024} and calibrate $\lambda$ for a particular experiment, namely $N=5$ and $\epsilon=0$, in which we obtain $\lambda=1.3$. The equilibrium configurations obtained in this case are shown in Fig. \ref{fig:magnet-simulations}, considering the $N=5$ and $N=10$ cases. Remarkably, we can reproduce the experimental observations very well regardless of the eccentricity using the same calibrated $\lambda$. It is expected, since this parameter depends solely on the balance between interparticle repulsion and confinement.

%The value of $\lambda$ in Eq. \eqref{eq:floaters_total_energy} is an adjustable parameter which can be calibrated using the experimental data. We address that using the GSA results with $N=5$ and $\epsilon=0$, we can reproduce the experimental observations very well with $\lambda = 1.3$. It is expected, since $\lambda$ depends solely on the balance between interparticle repulsion and confinement. The value of $\lambda$ is consistent with our expectations based on the dimensions of the floaters and the confining magnets, and does not change when $\epsilon$ changes. We also define in our simulations a cut-off distance $d/R = 0.01$ to prevent divergences in our simulations when two floating magnets are very close to one another or a floater approaches the bucket wall.

%is the balance between interparticle repulsion and confinement and $\hat{\mathbf{r}}_i = \mathbf{r}_{i}/a$ are the dimensionless position.  We stress that the coordinates of the latter vector are constrained to obey $\hat{x}_i^2 + \hat{y}_i^2\leq 1$.

%Regarding the confining field geometry, the fixed magnets are arranged to form an elliptical ring of eccentricity $e$, where we deliberately select the semi-axes as $a = (1 - e^2)^{-1/4}$ and $b = (1 - e^2)^{1/4}$ to honour the same cross sectional area used in the preceding experiments. This choice allows us to compare elliptical geometries with different eccentricities but having the same area.

\section{Comparison of the different systems}
\label{s:comparison}

%\remSH{I now think there should NOT be two subsections, experiments, simulations. The types of structures that we find in the exps are the same as in the sims. It all gets too much split up if we keep the results separate. I will work more on the text below to improve it.}

In Sect. \ref{s:experiments} we have presented experimental data for three different systems: hard spheres, floating magnets, and bubbles. All consist of a discrete set of interacting particles, confined by an external potential. These ingredients are sufficient to result in the same topological arrangements, despite the obvious differences in the character and nature of the respective interactions between particles. All the experimentally produced configurations are confirmed by simulations of the corresponding system, see Figs. \ref{fig:foam_sim} and \ref{fig:magnet-simulations}.

Hard sphere interactions are discontinuous and only act upon contact; they are usually modelled using a pairwise additive Hertzian potential~\cite{Duran2000}. Bubbles also interact via contact forces. However, in contrast to hard spheres, the change in shape which they undergo upon contact results in non-local forces, featuring a logarithmic term~\cite{HoehlerWeaire2019}. The floating magnets, on the other hand, are not governed by contact forces, but by long-range dipole interaction.

We studied arrangements of $N=5$ and $N=10$ particles of the above kinds, placed in an elliptic confining potential. The arrangements reveal topologically similar sequences of structures for the five different values of eccentricity, $\epsilon$, that we probed. In the following we will summarise these findings.

For $N=5$ and $\epsilon=0$ the centers of the particles (hard spheres/floating magnets/bubbles) can be regarded as the vertices of a (regular) pentagon (as in Fig.~\ref{fig:schematic-diagram}). The arrangement becomes increasingly distorted for larger values of $\epsilon$, leading to the loss of contacts of nearest neighbours. In the limit $\epsilon \to 1$ the only possible structure is that of a straight line of particles (see Table \ref{t:comparison} for floating magnets, and Table \ref{t:exp-bubbles-alternatives} for bubbles); in the case of hard spheres or floating magnets this straight line formed by particle centers is preceded by a zig-zag arrangement (or `buckled' line), which straightens out with increasing $\epsilon$. 

For bubbles the zig-zag structure is replaced by an arrangement with three bubble centers situated along the major axis, and two on a line perpendicular to it, but not coinciding with the minor axis (see Table \ref{t:comparison}, $\epsilon=0.95$). The same arrangement can also be obtained for our floating magnets (Table \ref{t:exp-bubbles-alternatives}, $\epsilon = 0.87$), but in this case it corresponds to a higher energy configuration than the zig-zag structure. The arrangement was also observed by
Saint Jean and Guthmann~\cite{SaintJeanGuthmann2002} in their experiments with steel spheres under the influence of Coulomb forces, for similar values of $\epsilon$. Apolonario {\em et al.} encountered it in their simulations of particles interacting via a repulsive logarithmic potential~\cite{ApolinarioEtal2005}.
 
%\remSC{Shouldn't we postpone any discussion of simulation results until we have discussed the patterns that {\em we} find?}

%\remSH{I think since we are near the end of the paper here might be the point to try and make connections. But it's tricky.}

An alternative arrangement for $\epsilon=0$ is to have one particle in the centre, surrounded by four particles in contact\footnote{We infer contacts between spheres from the images, rather than detecting them directly.} with the frame (or close to it, in the case of the floating magnets). This is shown in Appendix \ref{a:further-structures}, Table~\ref{t:exp-bubbles-alternatives}, for bubbles, floating magnets, and hard spheres. 
The corresponding 2D packing fraction of circles, $\phi=5/9$, is lower than that for the case when all circles are in contact with the frame, with $\phi_5=0.685$ \cite{packomania} (see section \ref{ss:hard-spheres}). Also for 2D foams, this arrangement is not optimal; our simulations show that it is of higher energy than having all bubbles in contact with the frame. In our foam experiments we were able to produce the higher energy structure for values of $\epsilon=$0, 0.44, and 0.66 (see Table ~\ref{t:exp-bubbles-alternatives}). In the case of floating magnets, the configuration was only stable at $\epsilon=0$. We associate the larger stability range found for foams with the fact that they constitute a `jammed system', requiring larger forces for a possible restructuring.

For $N=10$, a larger number of possible arrangements exist, see Tables \ref{t:comparison} and \ref{t:exp-bubbles-alternatives}; also here a similarity in the sequence of the `optimal' structures for different values of $\epsilon$ is apparent. Let us first turn to our sphere packings, which we will again interpret as a packing of circles. The densest packing of ten circles within a circle contains two circles in the centre, surrounded by eight circles in contact with the frame~\cite{packomania}. 
Each of the two central circles is in contact with three circles. Two of the perimeter circles contact three circles each, the remaining six have two contacts each. 
In our experiments with hard spheres this configuration is found also for $\epsilon=0.44$ and 0.66. 

The foam structure of minimal energy resembles the hard sphere arrangement in the following sense. It too consist consists of two particles (bubbles) in the centre, which are surrounded by eight bubbles which contact the frame. Also here two types of boundary particles can be identified. There are 6 bubbles contacting three bubbles each, and 2 bubbles contacting four bubbles each. Simulations show that this arrangement is that of lowest energy for $\epsilon < 0.84$, in our experiments we found it for $\epsilon = 0$, 0.44, and 0.66.

Also the arrangement of floating magnets (in both experiments and simulations) features two central and eight boundary magnets for $\epsilon = 0$, 0.44, and 0.66. In experiments using charged spheres the same arrangement was reported for $\epsilon=0.71$ \cite{SaintJeanGuthmann2002}.

At $\epsilon=0.87$ the structure differs from that at lower $\epsilon$ in all three systems.
The hard sphere structure still features two spheres that are not in contact with the frame, but each has gained an additional contact. The contact distribution of the eight boundary spheres is unchanged (6 $\times$ 2 contacts, and 2 $\times$ 3 contacts). Also the foam has still two bubbles in the centre, but each lost a side and now only contacts five boundary bubbles. Of these there are still two types, six with  four bubble contacts, and two with only two contacts. The magnet configuration is probably best described as four staggered rows of two magnets, each aligned perpendicular to the major axis, and one magnet each close to the endpoints of the major axis. Two of the magnets located towards the centre of the ellipse are furthest away from the frame and in this sense might be defined as the two non-boundary particles that are seen in both foams and hard sphere arrangements.

At $\epsilon=0.95$ all bubbles and all hard spheres are in contact with the frame, and there are also no more boundary magnets, in the sense defined above. The configurations in the three different systems all have $180^\circ$ rotational symmetry.

This type of arrangement is reminiscent of a line of hard spheres, placed in a transverse potential, which has buckled under compression \cite{HutzlerEtal2023, Ryan-PurcellEtal2025}. It is tempting to pursue this analogy, in order to explore what structural changes this would imply for this `zipper structure' \cite{Ryan-PurcellEtal2025} at an even higher value of $\epsilon$, corresponding to a {\em weaker}  compression along the major axis. As particle contacts gradually disappear (the zipper `unzips'), starting from the endpoints of the major axis (where there is no more room for pairs of particles), one arrives at a constellation, called the `doublet' \cite{Ryan-PurcellEtal2025}, in which all particles are aligned, except one pair close to the centre. A corresponding foam configuration is that for $N=5$ at $\epsilon=0.95$ in Table \ref{t:comparison}, a corresponding magnet configuration is shown in Table \ref{t:exp-bubbles-alternatives} for $N=5$ and $\epsilon=0.87$. 
The suggested analogy to the buckling experiments by \cite{HutzlerEtal2023, Ryan-PurcellEtal2025} also suggests the possibility of a `modulated zig-zag structure' \cite{WeaireEtal2022} in which the deviation of particle centres from the major-axis varies along the axis, with a maximum near the centre, before at even higher values of $\epsilon$ (i.e. lower compression) a straight line is formed. We indeed found such a modulated zig-zag formation in some preliminary experiments with a larger number of magnetic floaters. This could open up avenues for a theoretical analysis similar to those of Ref. \cite{WeaireEtal2022} in terms of a continuum model for structure formation.

%\remSC{And/or to the ordered structures by Weaire, Hutzler and Pittet?} 

\section{Discussion and Conclusion}
\label{s:discussion}

Based on the guiding model of Section \ref{energetics} we have seen that mutually repelling particles in the presence of confining forces self-assemble into clusters with topologies that are primarily determined by the confinement-repulsion ratio.  More specifically, in the case of circularly symmetric confinements, we have seen that it is possible to generate clusters with identical topologies even though the nature and range of the underlying interactions may be totally different. In addition, the same is possible in the case of elliptically symmetric confining forces, the main difference being that the eccentricity of the confining ellipse may be adjusted, together with the ratio of confining and repelling forces, in order to establish the exact cluster geometry. These findings suggest that it is possible to engineer cluster geometries whose topologies may display certain desirable functionalities~\cite{mark2024, blumler2025}. Bearing in mind that clusters of magnetic particles are crucial in many fields, including targeted drug delivery \cite{kianfar2021}, microwave absorption \cite{coatings11060621}, quantum computing \cite{doi:10.1021/jacs.7b12170}, molecular magnets and magneto-optical technologies \cite{doi:10.1021/ja209752z, doi:10.1021/ja5098622, doi:10.1021/acs.jpclett.3c03637}, being able to engineer the precise topology of such structures will only widen the range of applications for these clusters.

In the experiments that we presented, particles were placed into an elliptical frame of given eccentricity, $\epsilon$, and we recorded the resulting structure. In an alternative procedure one could vary $\epsilon$ continuously, keeping the area of the ellipse constant, with the particles remaining  confined. (This would not be possible for the hard sphere packings as their packing fraction varies with $\epsilon$.) At some critical values of eccentricity, rearrangements would occur. We would also expect hysteresis, i.e. different critical values, and even structures, depending on whether $\epsilon$ is increased or decreased. Experiments of this form might be difficult to devise, but simulations are possible. Indeed, the foam simulations shown in Fig.~\ref{fig:foam_sim} were already performed by varying $\epsilon$ in small increments, for a fixed configuration. However, since we were only interested in global energy minima, we did not determine the ranges of stability for the different structures.

%\nmauro{We haven't explored in detail the search for global minima for a wider range of $N$ with different potentials, perimeter minimzation, or hard-sphere packing. It is nonetheless of interest, and would need to be done so that the sort of self-assembly applications described above \remSH{(?) mention of applications above still lacks some detail} could be predicted.} 
%The appearance of alphaevolve, from google, gives an idea about what might be possible with super computing power (section B.12 onwards): 
%\href{https://colab.research.google.com/github/google-deepmind/alphaevolve_results/blob/master/mathematical_results.ipynb}{\underline{link}}

In all of the above we have only considered the (static) equilibrium configurations. The experiments using the floating magnets offer the opportunity to also study the dynamics.  In particular some of the metastable configurations are associated with an often minute-long ``dance'' of the floaters before they settle into equilibrium. Once this is achieved, individual floaters can be manually displaced to probe the impact this shows on the other floaters. The rate of relaxation towards the old or a new equilibrium configuration indicates the steepness of the energy gradient. In our experiments with floating magnets we observed `marginally' stable  structures that took much longer than others to reach equilibrium. 
In addition, moving particular magnets can have a more significant impact on the positions of the other magnets, leading to the possibility to identify particles which have some freedom of movement and others that are ``locked" into position. This may be related to floppy modes in condensed matter.

%\remSH{These are interesting questions and they could feature in this section. I certainly  (None of these structures are shown.)}

%extend to 3D?

\section*{Acknowledgments}
M. S. Ferreira thanks the Federal University of Rio Grande do Norte (UFRN) in Brazil for their hospitality in early 2025, during which most of this research was conducted. This publication has emanated from research supported in part by a research grant from Science Foundation Ireland (SFI) under Grant No. SFI/12/RC2778\_P2. S. Hutzler thanks D. Weaire for continued discussions of research ideas.
S. J. Cox was partially supported by the Horizon 2020 Framework Programme for Research and Innovation, Grant Agreement number 101008140 EffectFact.

\bibliography{ourbibliography}

\appendix
\onecolumngrid
\section{Further particle arrangements}
\label{a:further-structures}

In addition to the particle arrangements shown in Table \ref{t:comparison}, a variety of alternative structures were found in our experiments, see Table \ref{t:exp-bubbles-alternatives}.  Based on the results of our computer simulations, Section \ref{s:experiments}, they can be identified as metastable.

\begin{table*}[!h]
    \centering
    \begin{tabular}{|c|c|c|c|c|c|}
\hline
$\epsilon$      & 0 & 0.44 & 0.66 & 0.87 & 0.95 \\
\hline

%N=5 & \includegraphics[width=0.1\textwidth, angle = 0]{figures/bubble-experiments/Individual foam photos/e=0/N=5.png} & \includegraphics[width=0.1\textwidth, angle = 0]{figures/bubble-experiments/Individual foam photos/e=0.44/N=5.png} & \includegraphics[width=0.1\textwidth, angle = 0]{figures/bubble-experiments/Individual foam photos/e=0.66/N=5.png}& \includegraphics[width=0.1\textwidth, angle = 90]{figures/bubble-experiments/Individual foam photos/e=0.87/N=5.png}& \includegraphics[width=0.1\textwidth, angle = 90]{figures/bubble-experiments/Individual foam photos/e=0.95/N=5 draft.png}\\
%\hline

%N=5 & \includegraphics[width=0.1\textwidth, angle = 0]{figures/bubble-experiments/Individual foam photos/e=0/N=5.png} & \includegraphics[width=0.1\textwidth, angle = 0]{figures/bubble-experiments/Individual foam photos/e=0.44/N=5.png} & \includegraphics[width=0.1\textwidth, angle = 0]{figures/bubble-experiments/Individual foam photos/e=0.66/N=5.png}& \includegraphics[width=0.1\textwidth, angle = 90]{figures/bubble-experiments/Individual foam photos/e=0.87/N=5.png}& \includegraphics[width=0.1\textwidth, angle = 90]{figures/bubble-experiments/Individual foam photos/e=0.95/N=5 draft.png}\\

{\bf N=5} &  &  & & & \\
\begin{turn}{90} hard spheres \end{turn} & 
\includegraphics[width=0.13\linewidth, angle = 0]{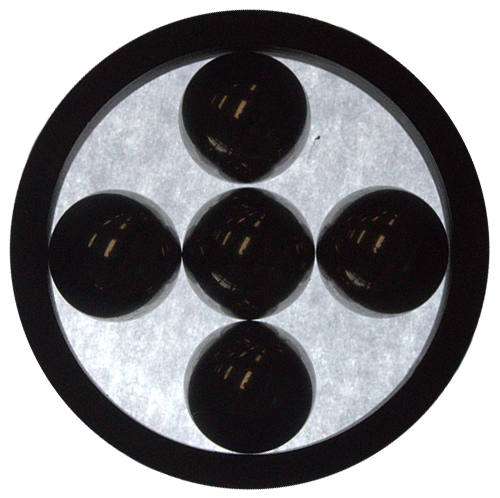} 
&  &  & ~ & ~ \\
~ & only $\epsilon=0$ was set up & & & & \\
~ & for this topology &  & & & \\

\hline

\begin{turn}{90} floating magnets \end{turn} & 
\includegraphics[width=0.13\linewidth, angle = 0]{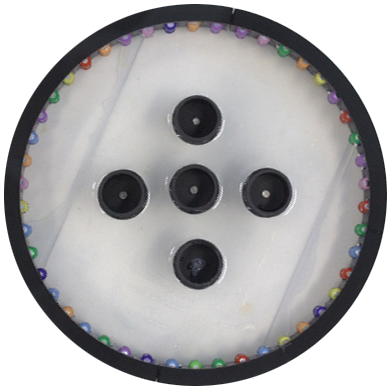}
& ~ & ~ &
\includegraphics[width=0.1851\linewidth, angle = 0]{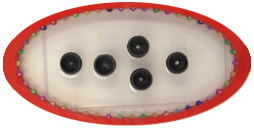}
& ~ \\
& only stable for $\epsilon=0$ & & & &\\

\hline

\begin{turn}{90} bubbles \end{turn} & 
\includegraphics[width=0.13\linewidth, angle = 0]{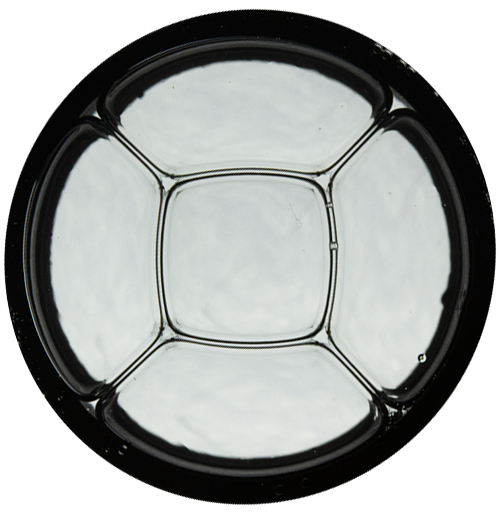}
&
\includegraphics[width=0.1372\linewidth, angle = 0]{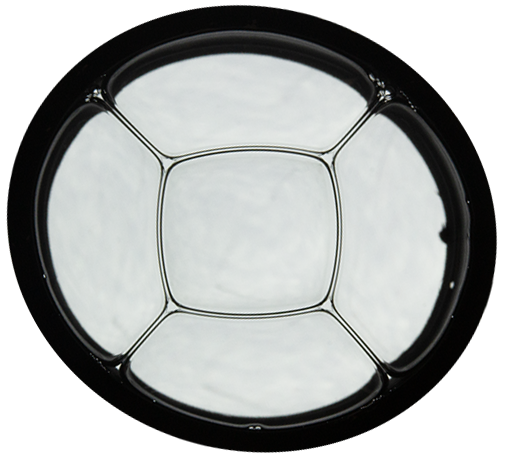}
& 
\includegraphics[width=0.1500\linewidth, angle = 0]{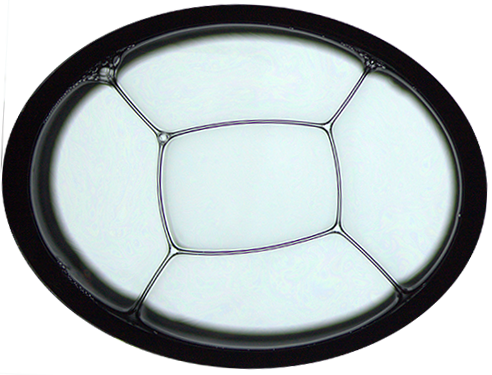}
& 
& 
\includegraphics[width=0.2327\linewidth, angle = 0]{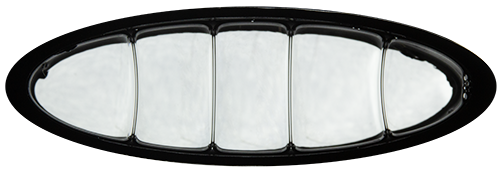}
\\

\hline
\hline

%N=10     & \includegraphics[width=0.1\textwidth, angle = -10]{figures/bubble-experiments/Individual foam photos/e=0/N=10.png} &  \includegraphics[width=0.1\textwidth, angle = 90]{figures/bubble-experiments/Individual foam photos/e=0.44/N=10.png}& \includegraphics[width=0.1\textwidth, angle = 0]{figures/bubble-experiments/Individual foam photos/e=0.66/N=10.png}& \includegraphics[width=0.1\textwidth, angle = 90]{figures/bubble-experiments/Individual foam photos/e=0.87/N=10.png}& \includegraphics[width=0.1\textwidth, angle = 90]{figures/bubble-experiments/Individual foam photos/e=0.95/N=10.png} \\
%\hline

%N=10     & \includegraphics[width=0.1\textwidth, angle = 90]{figures/bubble-experiments/Individual foam photos/e=0/N=10.png} &  \includegraphics[width=0.1\textwidth, angle = 90]{figures/bubble-experiments/Individual foam photos/e=0.44/N=10.png}& \includegraphics[width=0.1\textwidth, angle = 0]{figures/bubble-experiments/Individual foam photos/e=0.66/N=10.png}& \includegraphics[width=0.1\textwidth, angle = 90]{figures/bubble-experiments/Individual foam photos/e=0.87/10.2.png}& \includegraphics[width=0.1\textwidth, angle = 90]{figures/bubble-experiments/Individual foam photos/e=0.95/N=10.png} \\

{\bf N=10} & & & & & \\

\begin{turn}{90} floating magnets \end{turn} & 
\includegraphics[width=0.13\linewidth, angle = 0]{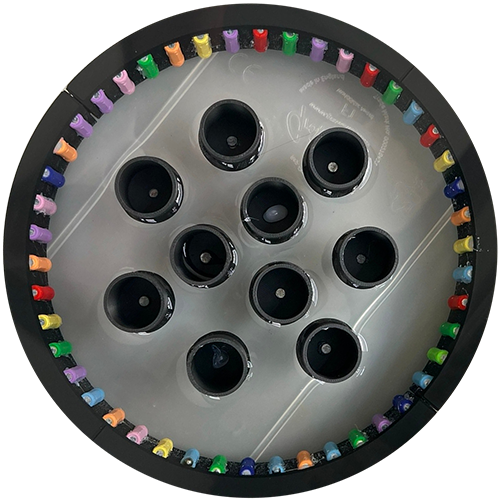}
& ~ & ~ &  & ~ \\
& only stable for $\epsilon=0$ & & & &\\

\hline

\begin{turn}{90} bubbles \end{turn} & 
\includegraphics[width=0.13\linewidth, angle = 0]{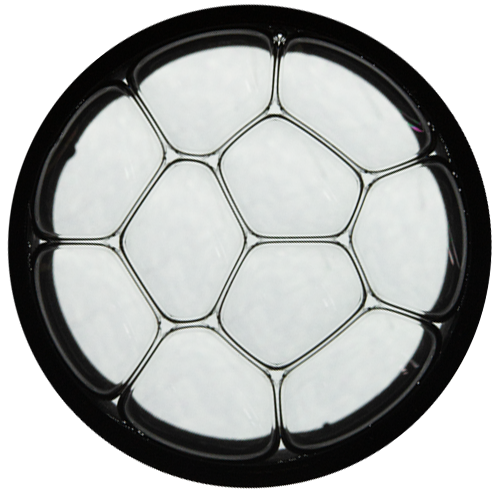}
& 
\includegraphics[width=0.1372\linewidth, angle = 0]{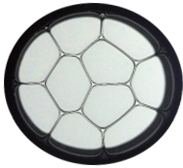}
&
\includegraphics[width=0.150\linewidth, angle = 0]{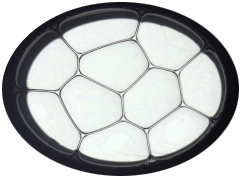}
&
\includegraphics[width=0.1851\linewidth, angle = 0]{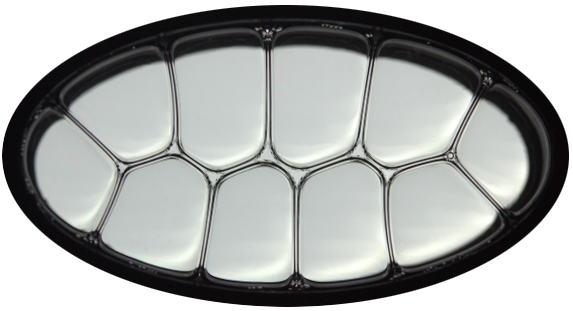}
&\\

&  &
&
\includegraphics[width=0.150\linewidth, angle = 0]{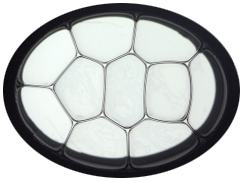}
&
\includegraphics[width=0.1851\linewidth, angle = 0]{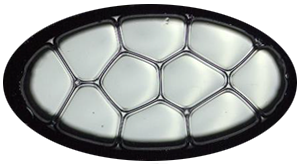}
&\\

\hline

\hline

\end{tabular}
\caption{
Some alternative arrangements obtained in experiments with 5 and 10 particle clusters in elliptical confinement, cf. Table \ref{t:comparison}.
}
\label{t:exp-bubbles-alternatives}
\end{table*}

\end{document}